# Informe Técnico / Technical Report

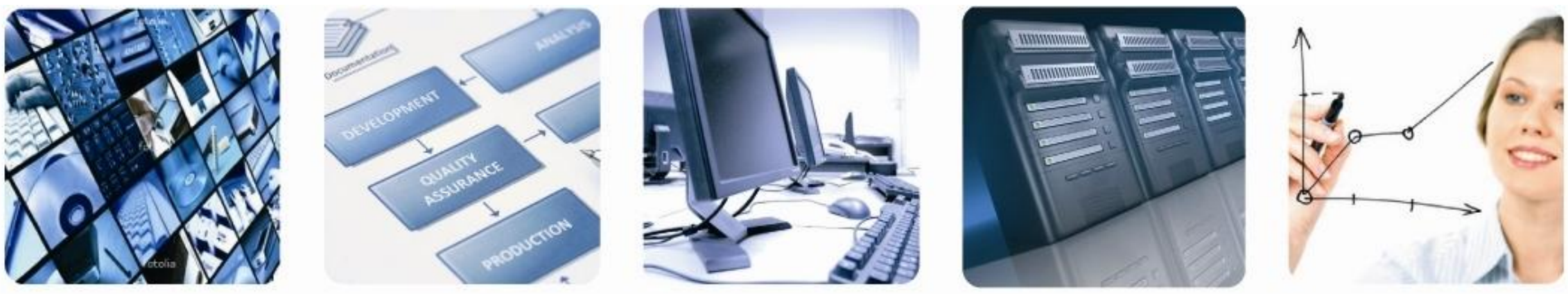

## BP Variability Case Studies Development using different Modeling Approaches

Clara Avora. Victoria Torres. Vicente Pelechano

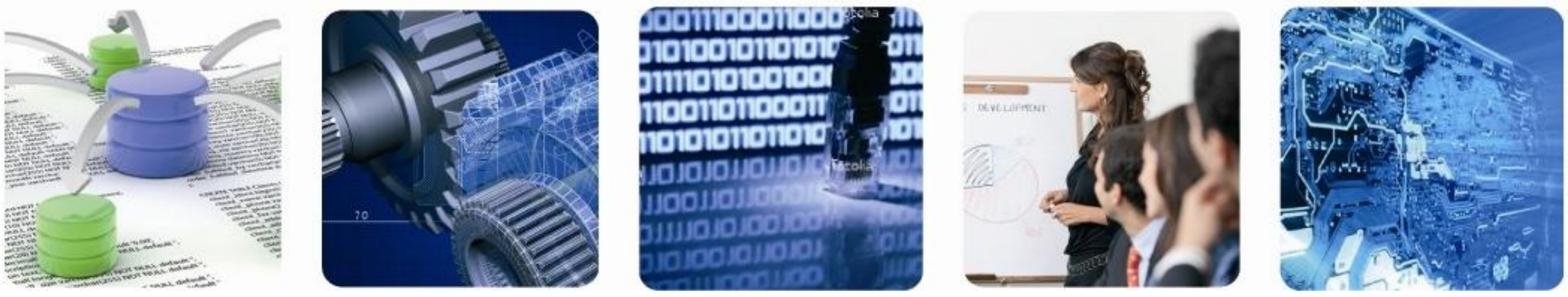

| | |
|---|---|
| Ref. #: | ProS-TR-2011-03 |
| Title: | BP Variability Case Studies Development using different Modeling Approaches |
| Author (s): | Clara Ayora, Victoria Torres and Vicente Pelechano |
| Corresponding author (s): | cayora@pros.upv.es<br>vtorres@pros.upv.es<br>pele@pros.upv.es |
| Document version number: | 1.2 · Final version: No · Pages: 30 |
| Release date: | February 2011 |
| Key words: | Business Process Modeling, Model Variability |

Universidad Politécnica de Valencia · Camino de Vera s/n · Edificio 1F · 46022 Valencia Spain ·
T. +34 96 387 70 07 Ext. 83530 · M. + 34 619 543 623 · F. +34 96 387 73 59 · info@pros.upv.es · www.pros.upv.es

# BP Variability Case Studies Development using different Modeling Approaches

Clara Ayora, Victoria Torres and Vicente Pelechano

Centro de Investigación en Métodos de Producción de Software
Universidad Politécnica de Valencia
Camino de Vera s/n, 46022, Valencia, Spain
{cayora, vtorres, pele}@pros.upv.es

# Index

# 1. Introduction

Variability in Business Process modeling has already been faced by different authors from the literature. Depending on the context in which each author faces the modeling problem, we find different approaches (C-EPC [1], C-YAWL [2], FEATURE-EPC [3], PESOA [4], PROVOP [5], or WORKLETS [6]). In this report we present four of the most representative approaches (C-EPC, PESOA, PROVOP and WORKLETS) which are presented by means of the different case studies found in the literature.

The remainder of this technical report is organized as follows. Next section presents each of the case studies used in this report and, for each of them, applies the four different approaches selected from literature.

# 2. Approaches Overview

- **Approach: Configurable Event-driven Process Chains (C-EPC)**

C-EPC is an extension of EPC (Event-driven Process Chain) that includes new constructs to represent variability in reference process models. The main idea of C-EPC is to represent differently commonalities from individualities in order to configure the process model according to its context. This differentiation is possible by combining the use of (1) *configurable nodes* (functions and connectors), which allow specifying different behavior depending on the context of use with (2) *configuration requirements and guidelines*, which state, by means of logical predicates, the valid configurations of the model.

The complete description of the approach is presented in [1].

- **Approach: Rich BPMN (PESOA)**

PESOA (www.pesoa.org) is a cooperative project carried out by a group of companies (DaimlerChrysler AG, Delta Software Technology GmbH, ehotel AG and Fraunhofer IESE) and academics from the Hasso-Plattner-Institute and the University of Leipzig whose main objective was the investigation of a proposal for the development and customization of families of process oriented software. As a result, focused at the design level and taking into account the relevance of the reusability aspect, a set of basic and composite variability mechanisms was identified. The basic set includes (1) Encapsulation of Varying Subprocesses, (2) Addition, Replacement, Omission of Encapsulated Subprocesses, (3) Parameterization, and (4) Variability in Data Types, while the composite include (5) Inheritance, (6) Design Patterns, and (7) Extensions/Extension Points. This set was transferred to different languages such as UML Activity Diagrams, UML State Machines, BPMN, and Matlab/Simulink. Since we are focused on the concepts and not in the particularities of any language, we have taken the transference made to BPMN to evaluate the PESOA proposal.

The complete description of the approach can be found in [4].

- **Approach: Process Variants by Options (PROVOP)**

Provop is an operational approach for managing large collections of process variants during the process life cycle. It has been motivated by the fact that a process variant can be created by adjusting (configuring) a given process model to a given context. These variant-specific adjustments are expressed by means of a set of high-level change operations (INSERT, DELETE, MOVE and MODIFY). Furthermore, Provop allows more complex process adjustments by grouping multiple change operations into so called *options*. Thus, a particular process variant is specified (configured) by applying one or more options to the respective base process. The options used for a process variants are selected when evaluating the given context. Provop provides a model for capturing this process context by means of *context variables*, which represent different domain dimensions of it.

A complete description of the approach is presented in [5].

- **Approach: Worklets (YAWL+RDR)**

This proposal is an approach for dynamic flexibility, evolution and exception handling in workflows through the support of flexible work practices. It was not conceived to be targeted to any notation (language independence) which means that it can be applied to any BPML. A *worklet* is defined as a small, complete and re-usable workflow specification which handles one specific task in a composite parent process. In this parent process, an *extensible repertoire* (or catalogue) of worklets is maintained for each nominated task. Each time a worklet is needed, an intelligently choice is made from this repertoire using a set of associated *selection rules* (Ripple Down Rules, RDR). These rules determine the most appropriate substitution. Then, the selected worklet is launched as a separate case and, when it has completed, the control is returned to the original (parent) process, which continues normally. Thus, dynamic ad-hoc change and process evolution are provided without having to modificate the original process specification and/or to resort to off-system intervention.

The full description of the approach can be found in [6].

# 3. Case Studies

This section presents a complete description of each case study followed by the different models obtained after applying each one of the evaluated approaches.

## Vehicle Repair Process

This case study is taken from the work developed by *Hallerbach et al.* in [7] to present their approach Provop. This case study is developed in the context of the automobile industry.

The process starts with the **reception** of a vehicle. After a **diagnosis** is made, the vehicle is **repaired** if necessary. During its diagnosis and repair the vehicle is **maintained**; e.g. oil and wiper fluid are checked and refilled if necessary. The process ends when **handing over** the repaired and maintained vehicle to the customer. Depending on the process context, different variants of this process are required, whereas the context is described by country-specific, garage-specific, and vehicle specific variables.

**Variant 1**: Assumes that the damaged vehicle requires a checklist of "Type2" to perform the diagnosis. Therefore, activities **diagnosis** and **repair** are adapted by modifying their attribute **checklist** to value "type2". Additionally, the garage omits maintenance of the vehicle as this is considered as special service not offered conjointly with the repair process.

**Variant 2**: Due to country-specific legal regulations, a final security check is required before handing over the vehicle back to the customer. Regarding this variant, new activity **final check** has to be added when compared to the standard process.

**Variant 3**: If a checklist of "type 2" is required for vehicle diagnosis and repair, the garage does not link maintenance to the repair process, and there are legal regulations requiring a final security check.

- **Approach: Configurable Event-driven Process Chains (C-EPC)**

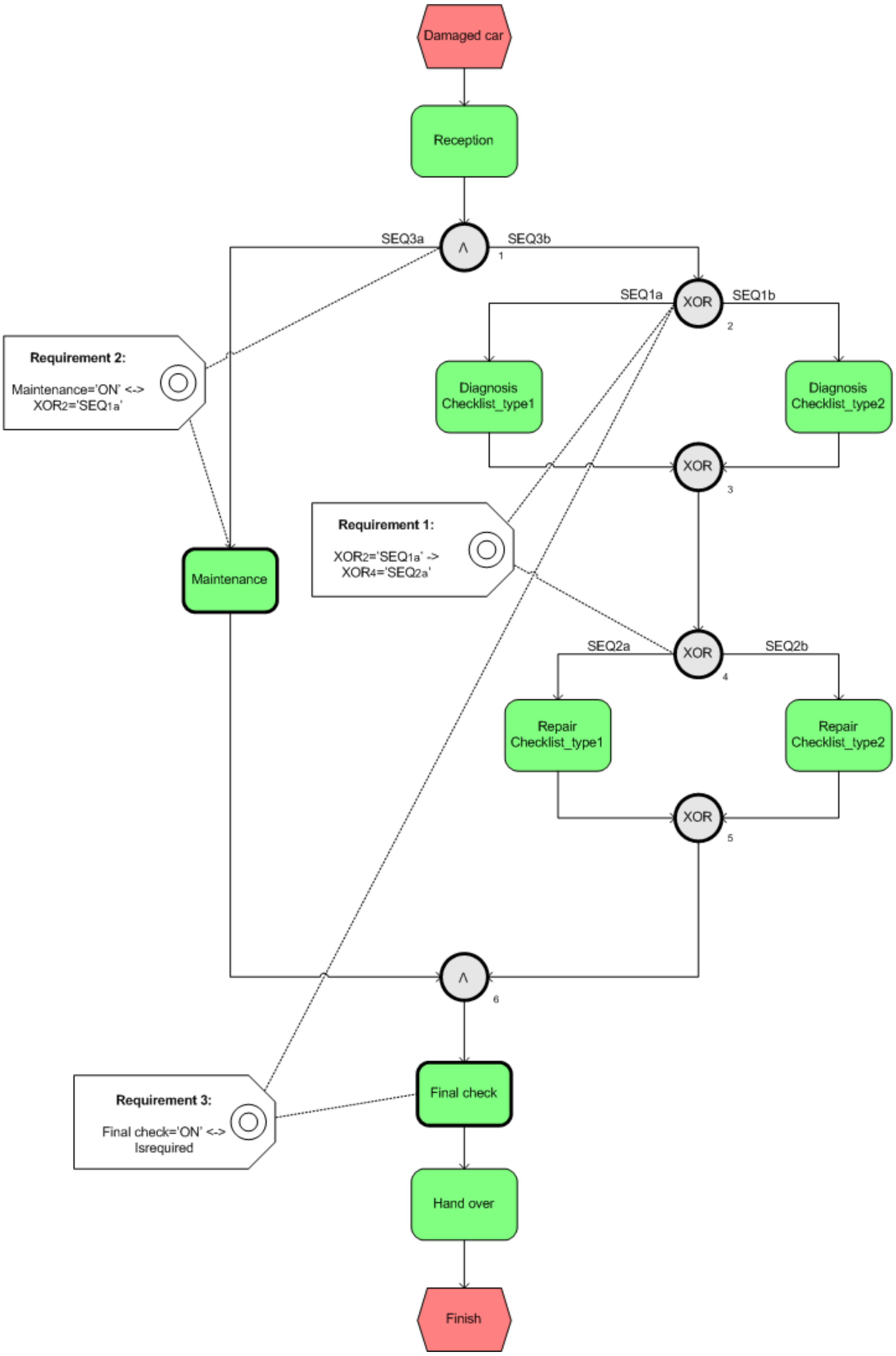

- **Approach: Rich BPMN (PESOA)**

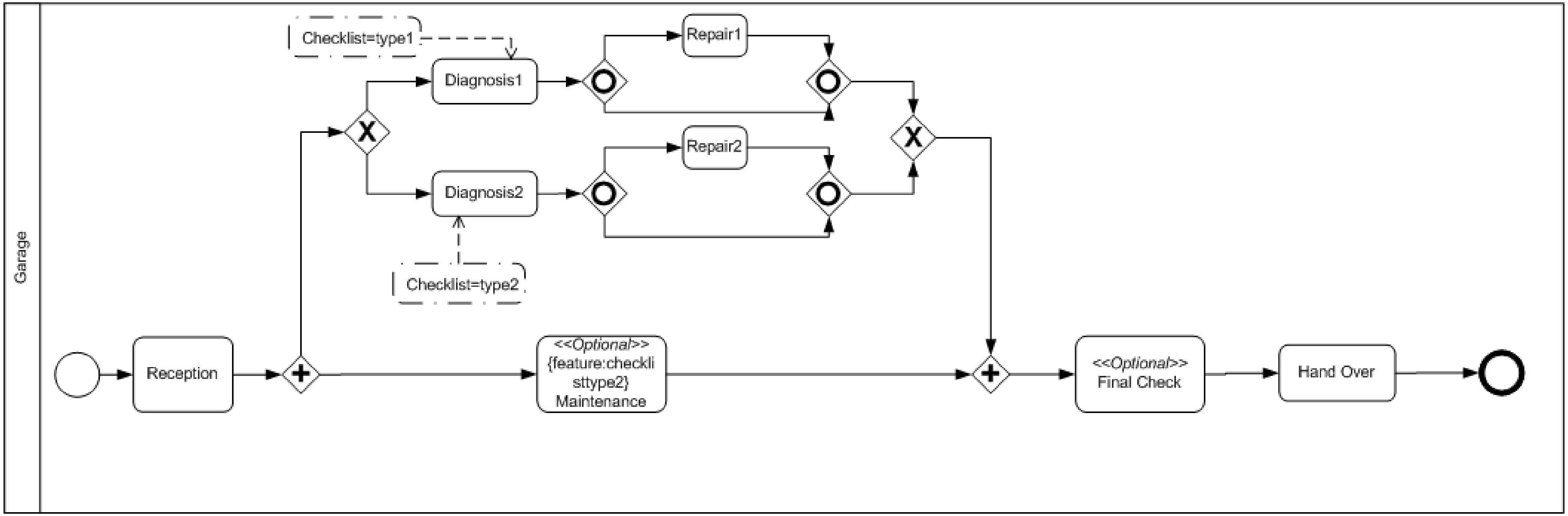

- **Approach: Process Variants by Options (PROVOP)**

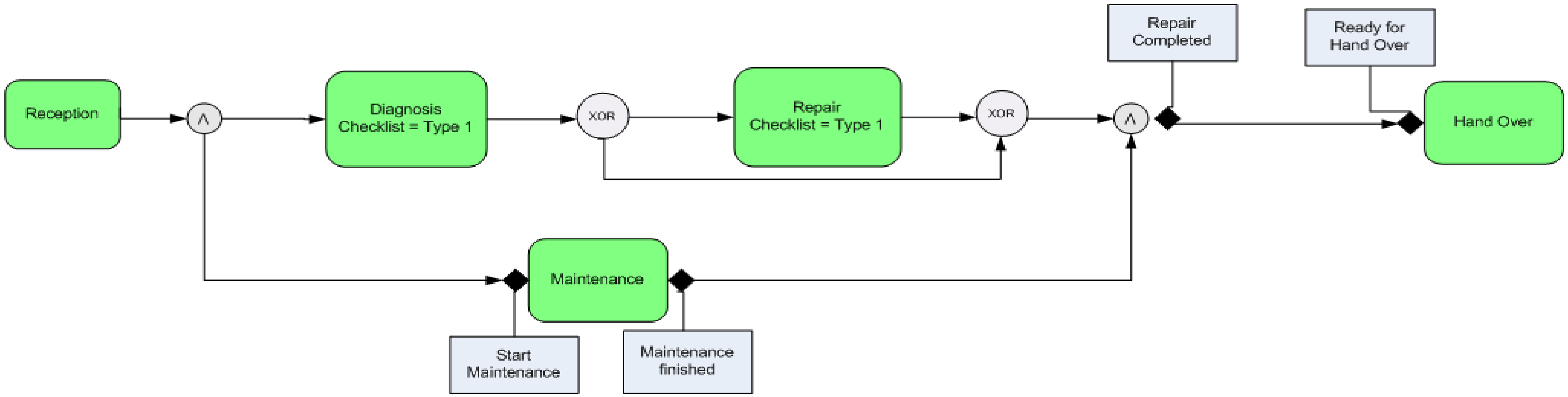

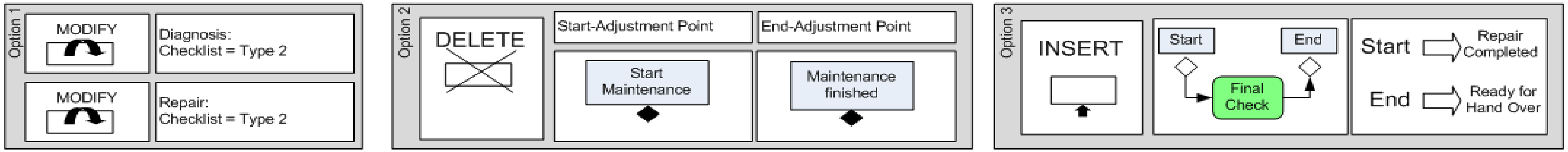

Variant 1: Options 1 and 2 (together)
Variant 2: Option 3
Variant 3: Option 2 and (then) Option 3

- **Approach: Worklets (Yawl+RDR)**

Worklet-Enabled process:

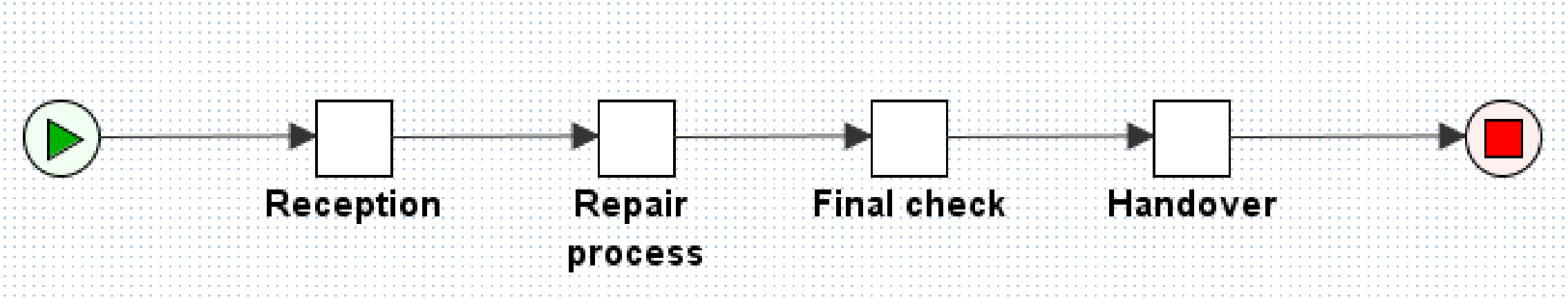

Selection Rule Tree for the “Repair process” task:

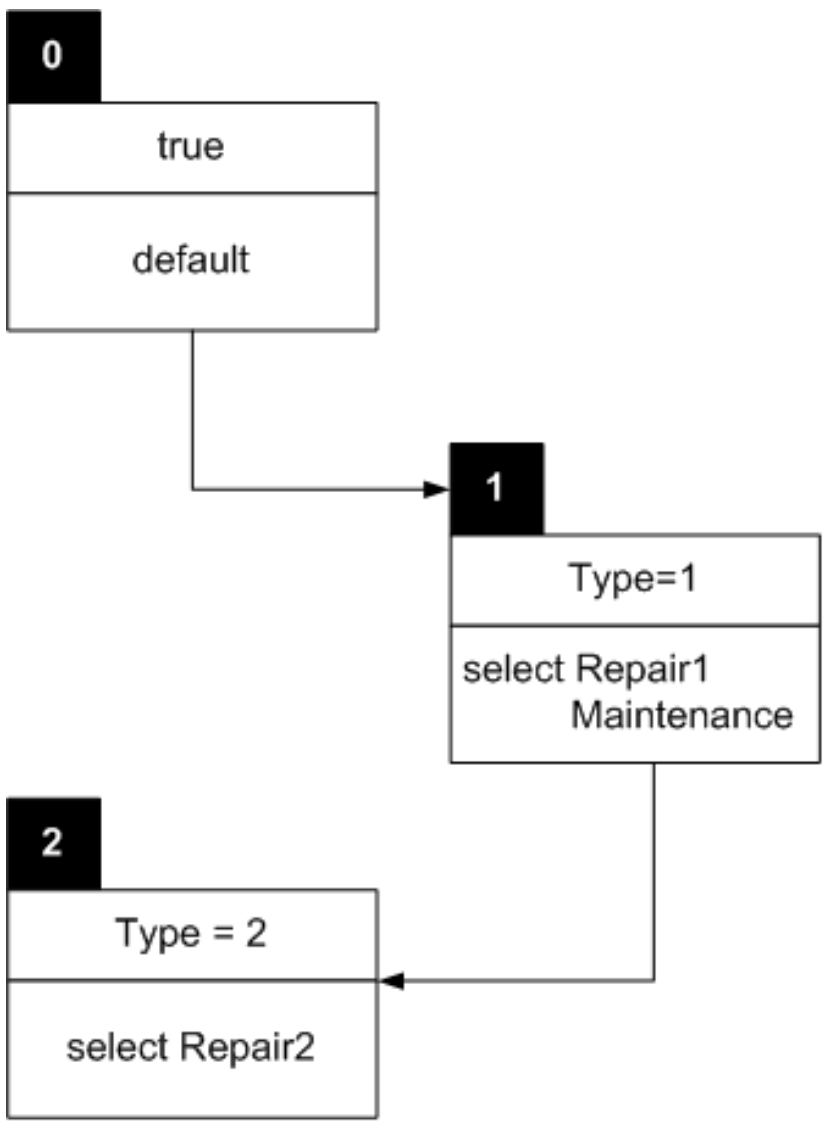

Selection Rule Tree for the “Final check” task:

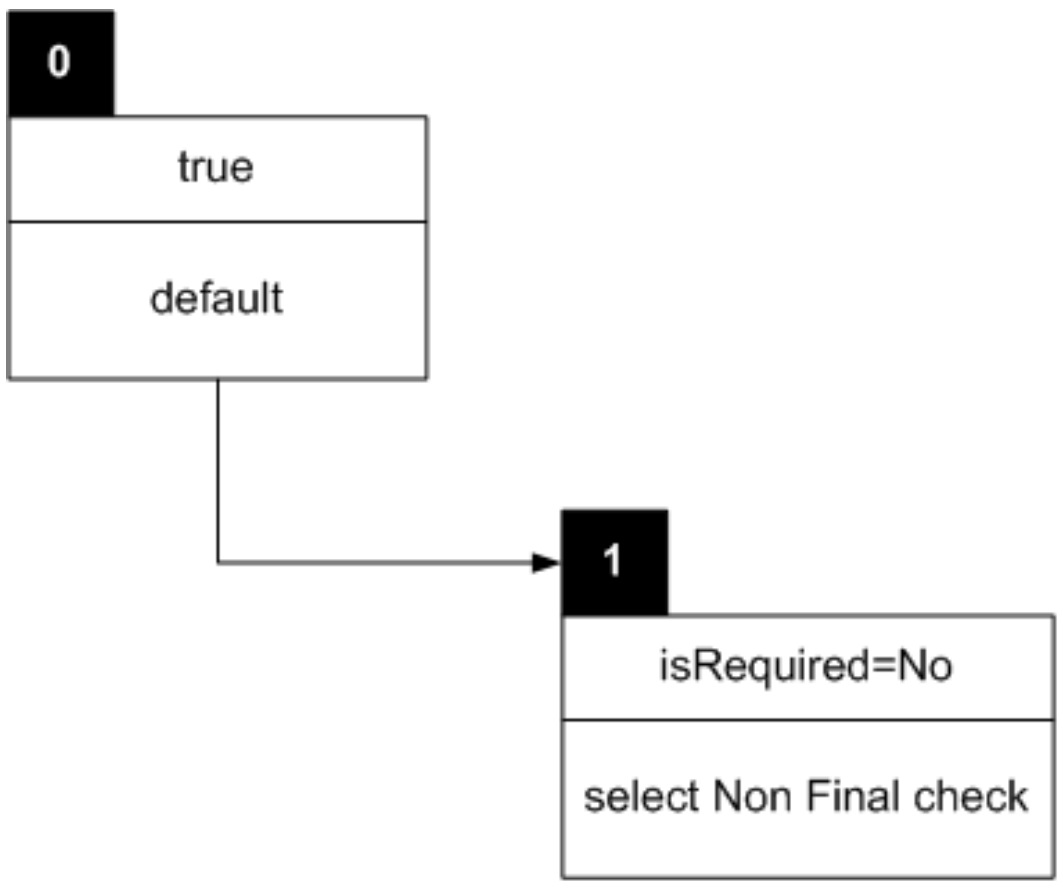

Worklet Repair1:

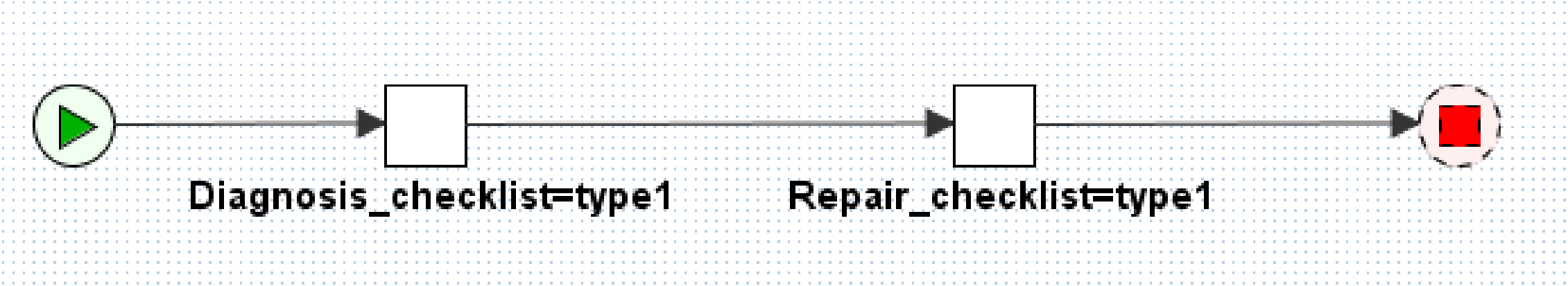

Worklet Repair2:

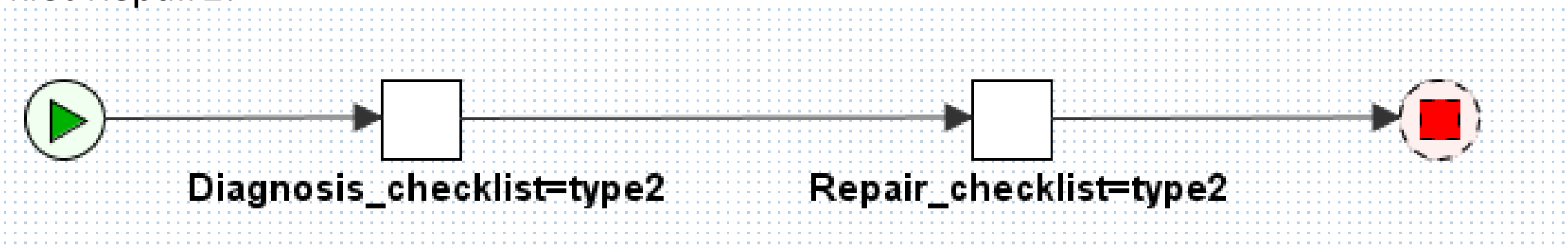

Worklet Maintenance:

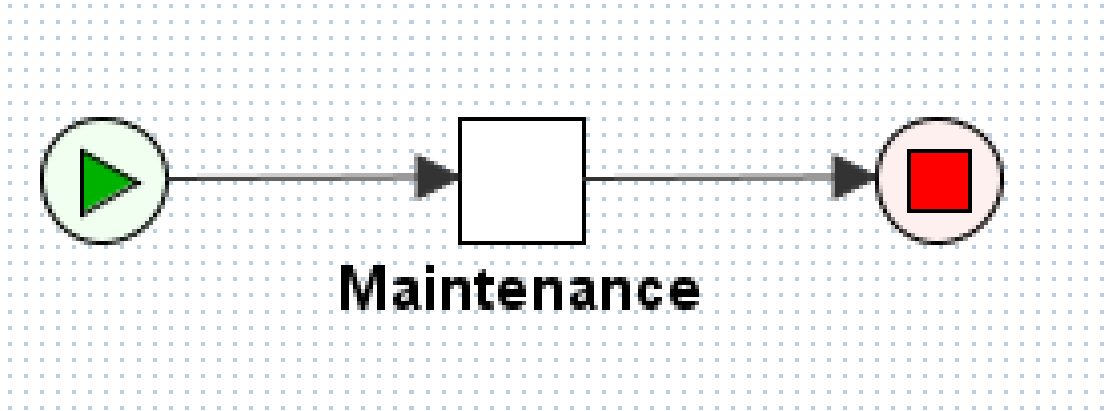

Worklet Non Final check:

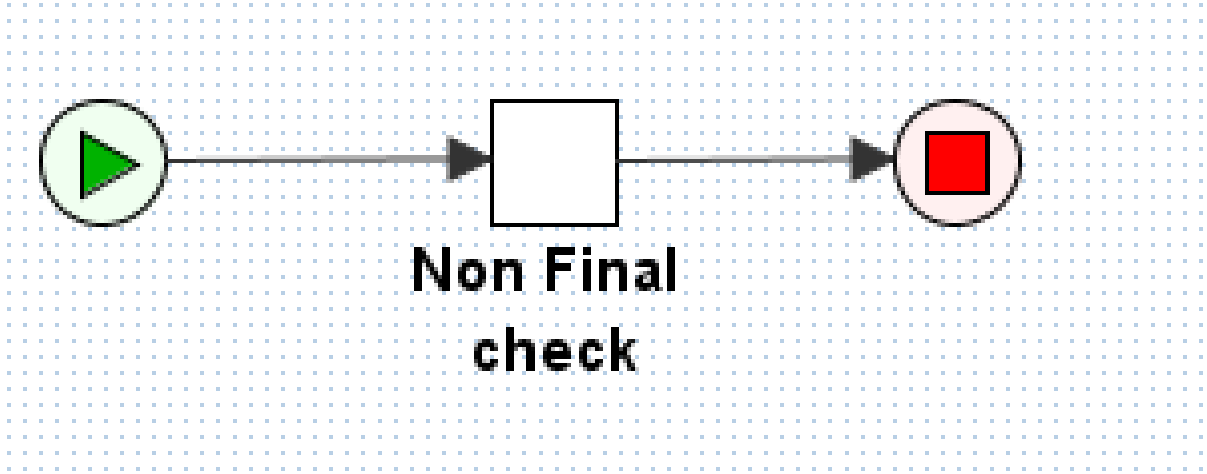

## Healthcare process

This case study is taken from [8]. It has being developed within the healthcare domain and shows a simplified version of a healthcare process representing a cruciate rupture treatment.  It is modeled in BPMN.

The process is started by the **Admission of the patient**. Then, the **Anamnesis and Clinical Examination** is performed to the patient. After this examination, different tests (**X-ray**, **MRT**, and **Sonography**) can be processed in parallel in any arbitrary order. Only in the case that the patient is suffering from a cruciate rupture the activities **Initial Treatment and Operation planning** and **Operative Treatment** will be performed.

**Variant 1**: When patients with cardiac pacemaker skip **MRT** test.

**Variant 2**: When patients suffer from an effusion in a knee, a **puncture** has to be done.

**Variant 3**: Due to legal regulations, it would be necessary to inform the patients about the treatment.

- **Approach: Configurable Event-driven Process Chains (C-EPC)**

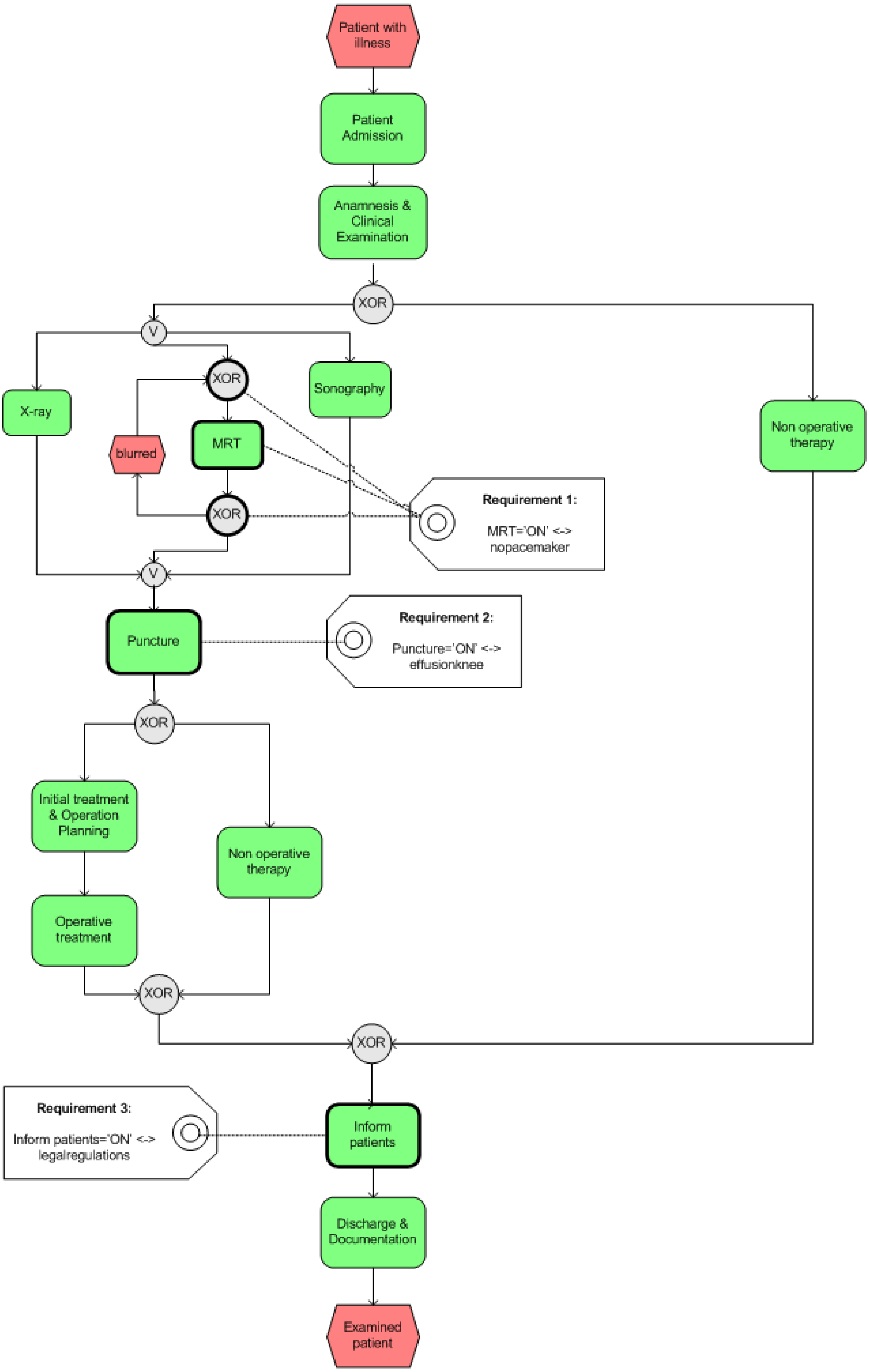

- **Approach: Rich BPMN (PESOA)**

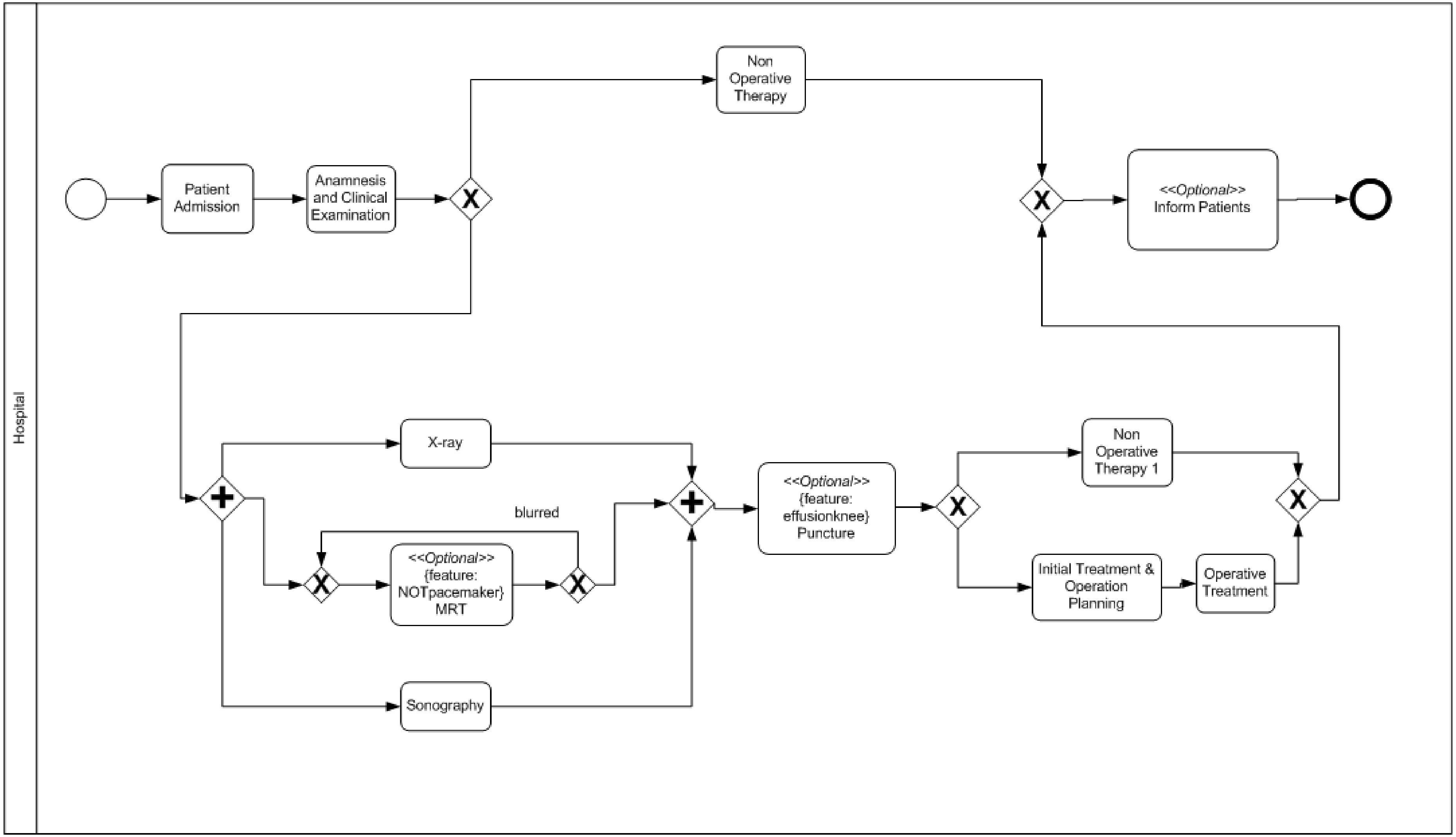

- **Approach: Process Variants by Options (PROVOP)**

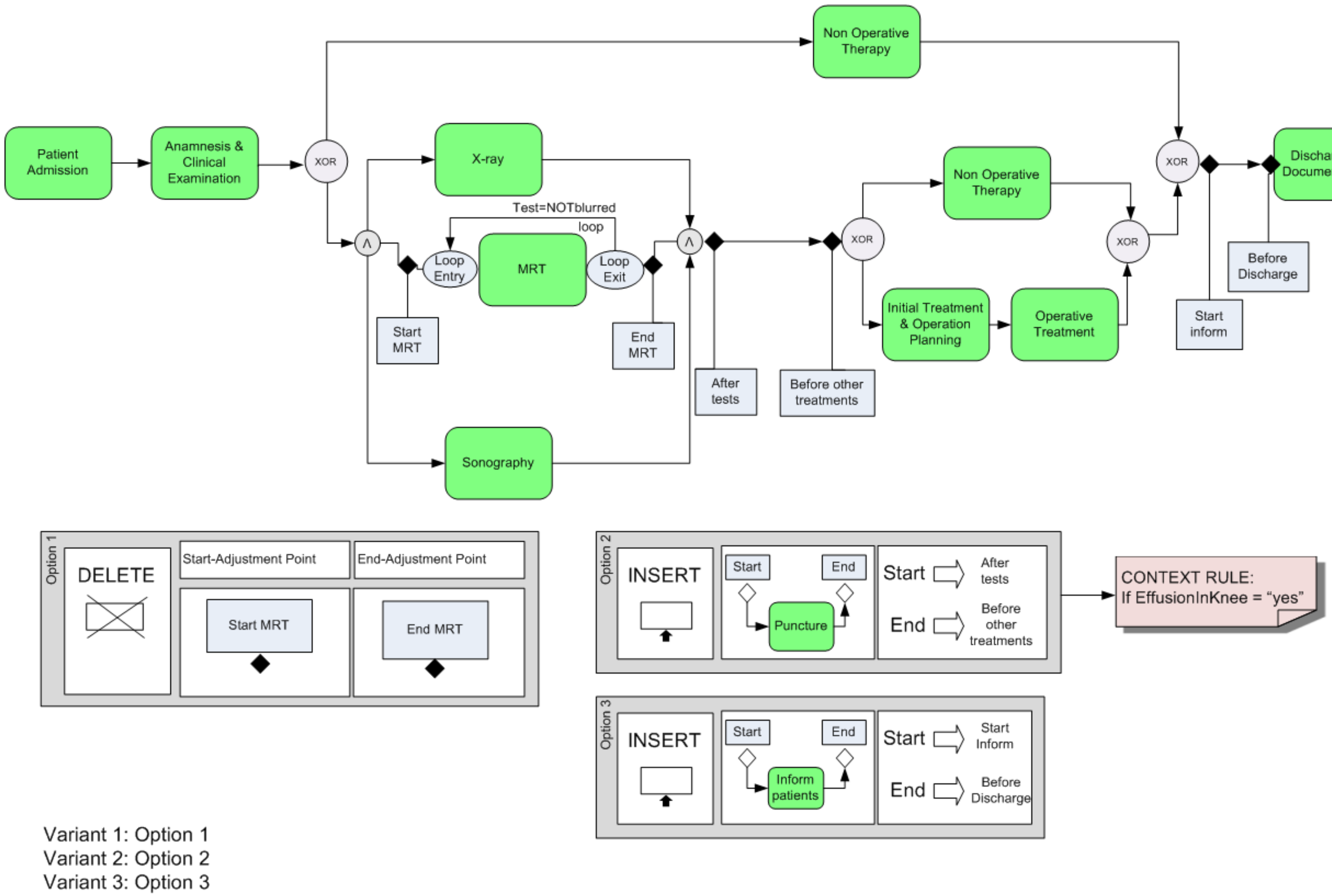

- **Approach: Worklets (Yawl+RDR)**

Worklet-Enabled process:

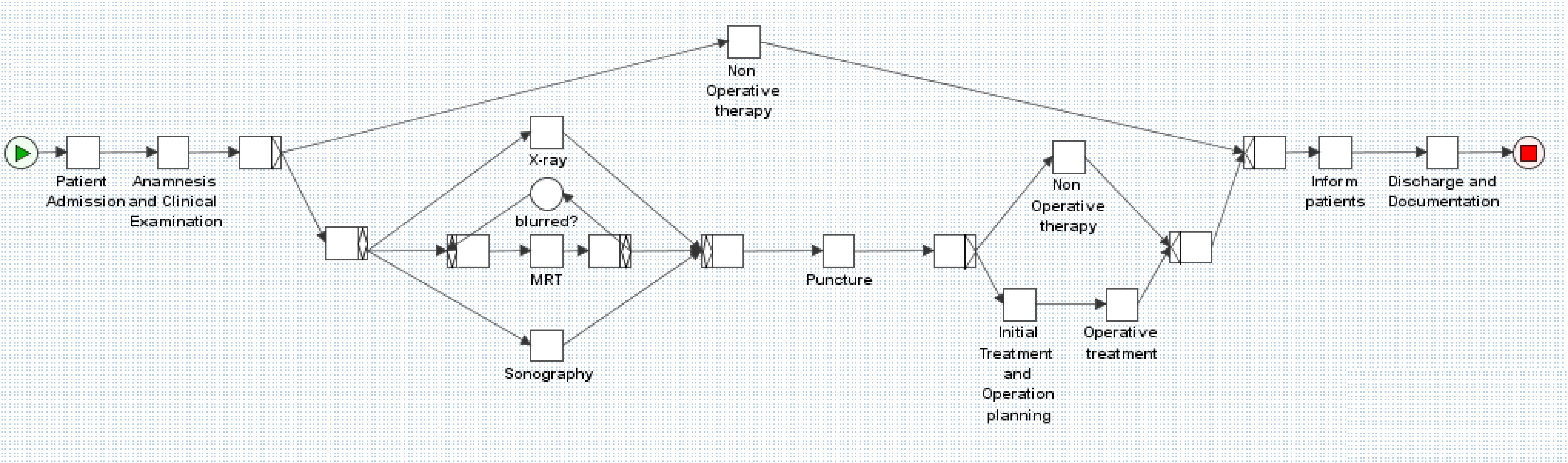

Selection Rule Tree for the “MRT” task:

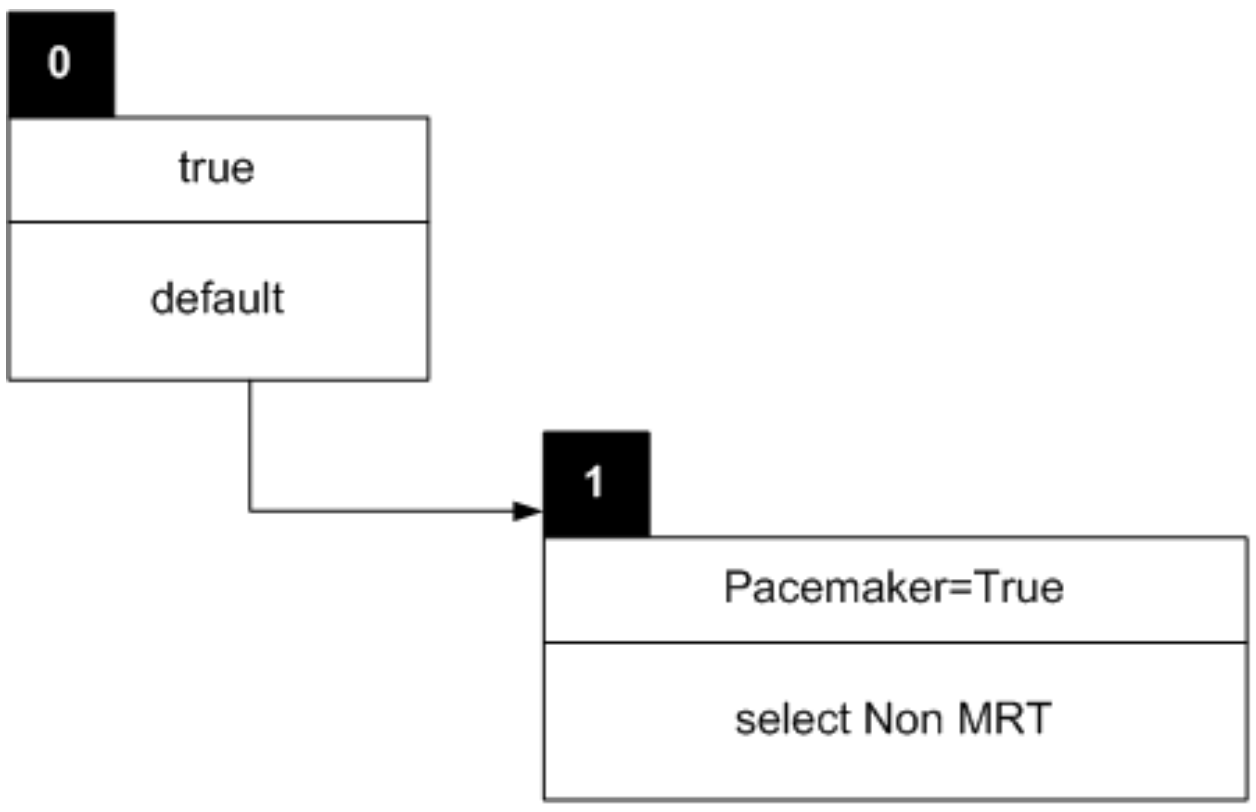

Selection Rule Tree for the “Puncture” task:

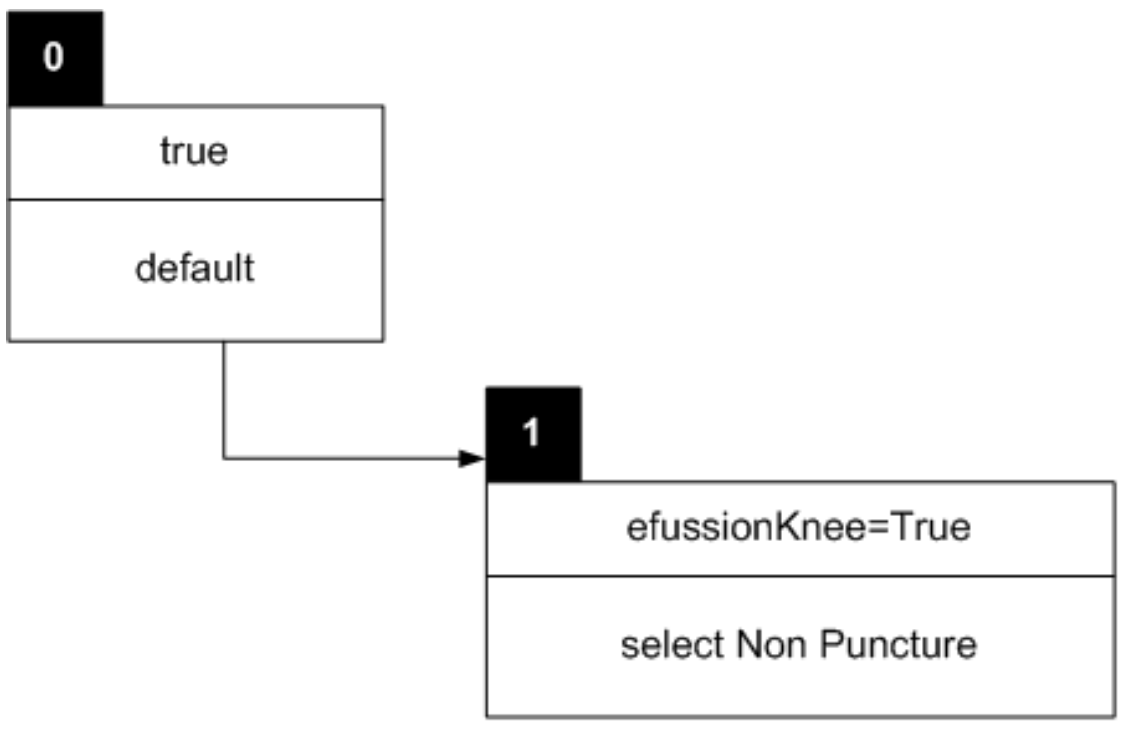

Selection Rule Tree for the “Inform patients” task:

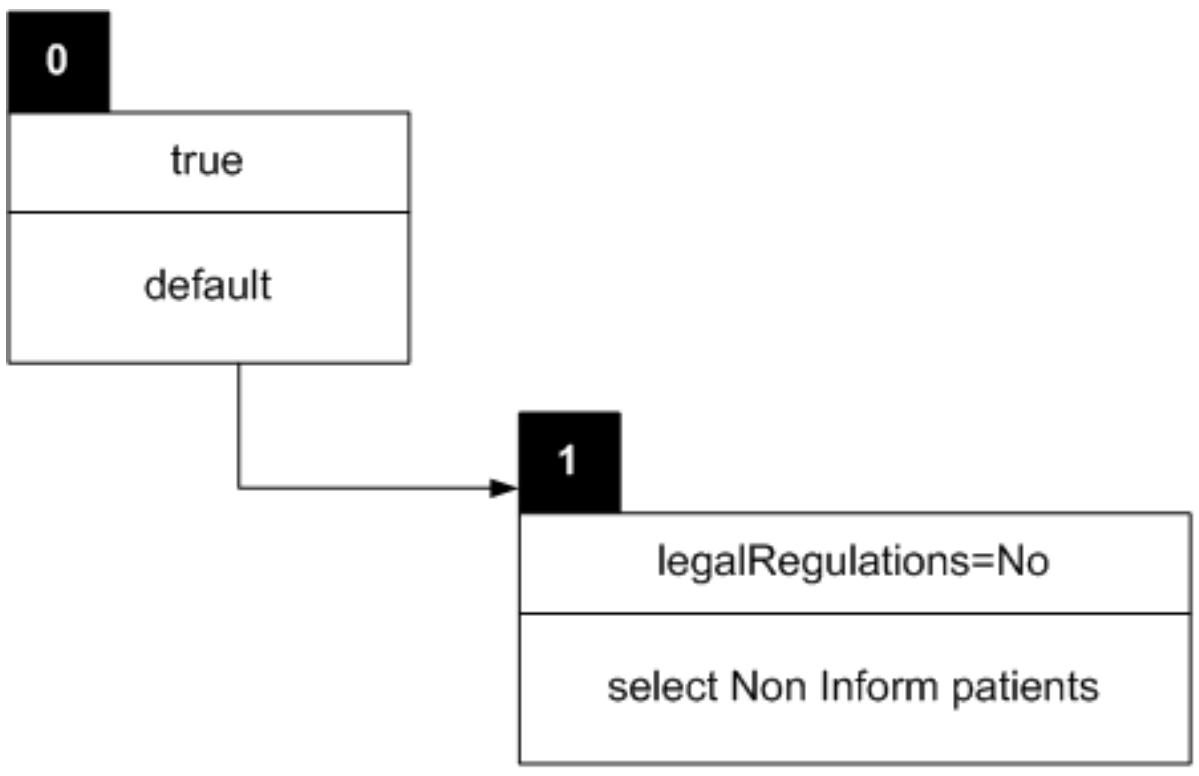

Worklet Non MRT:

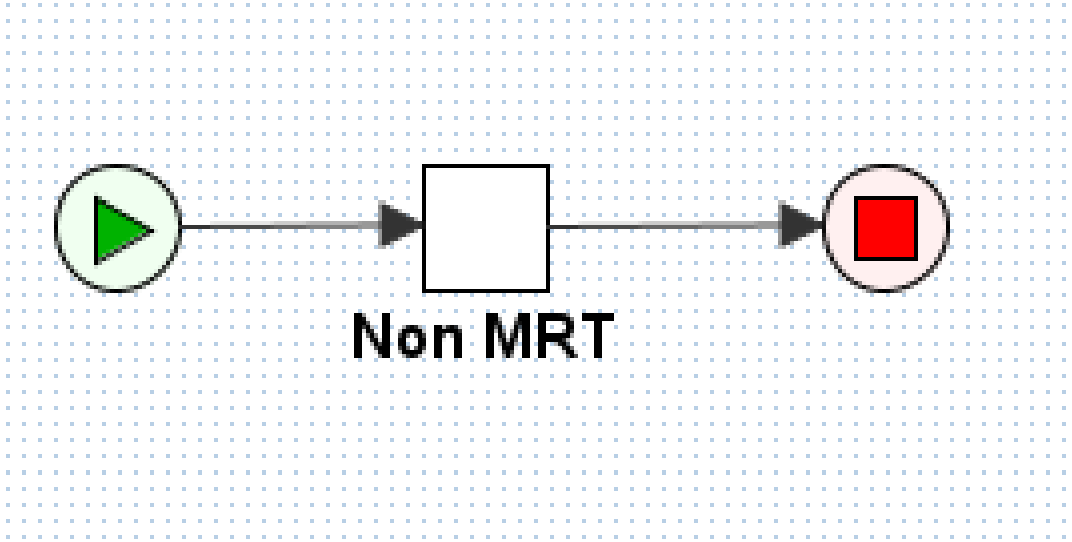

Worklet Non Puncture:

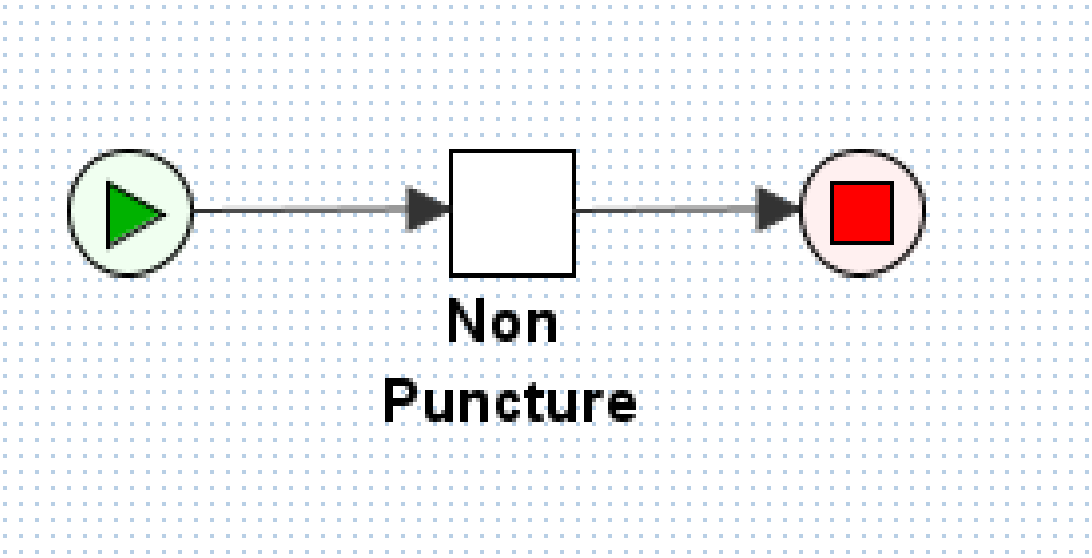

Worklet Non Inform patients:

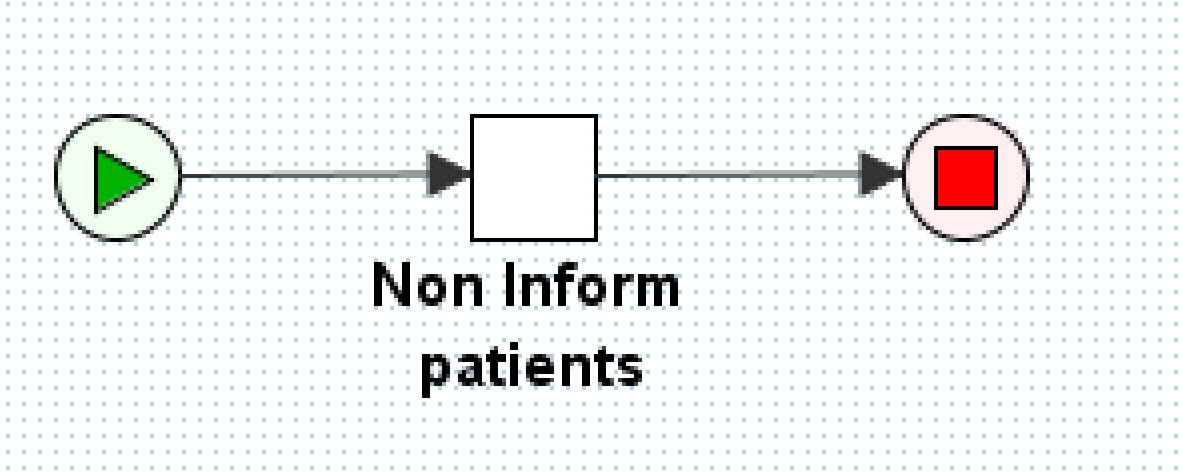

## e-business shop

This case study is taken from the PESOA report [9] developed in the context of the BMBF-Project. This case study is developed in the context of the e-commerce domain (B2C market) using the BPMN notation.

Two different roles interact: the *customer* and the *shop.* The process is started by the *customer* who **explores** the products that the shop offers to its clients. During this exploration, the *shop* **delivers** to the *customer* the appropriate information of the products. Then, once the *customer* has **chosen** some of the products, the *shop* **composes** the *customer* shopping-cart. During the next step, the *customer* decides to buy the selected products. At this point, the *shop* starts with the **checkout** process and then with the **delivery** of the selected products. If the *customer* takes a long time to perform the purchase, then the process is finished. If not, the products are **received** by the customer.

**Variant 1**: The information provided to the *customer* regarding the product consist of a textual description and optionally pictures and reviews.

**Variant 2**: The shopping cart can be made persistent optionally.

**Variant 3**: The shop could support personalized shopping carts. This means that the 10% of the purchase is given to the customer to buy what she/he wants. Another option is to have an anonymous shopping cart. In this case, the identity of the customer is hidden. The shop should only support one of these two types of shopping carts.

**Variant 4**: The checkout task should support the payment by credit card. However, if a personalized shopping cart is selected, then the invoice payment is offered optionally.

**Variant 5**: If the customer takes a long time to perform the purchase, then the process is finished.

- **Approach: Configurable Event-driven Process Chains (C-EPC)**

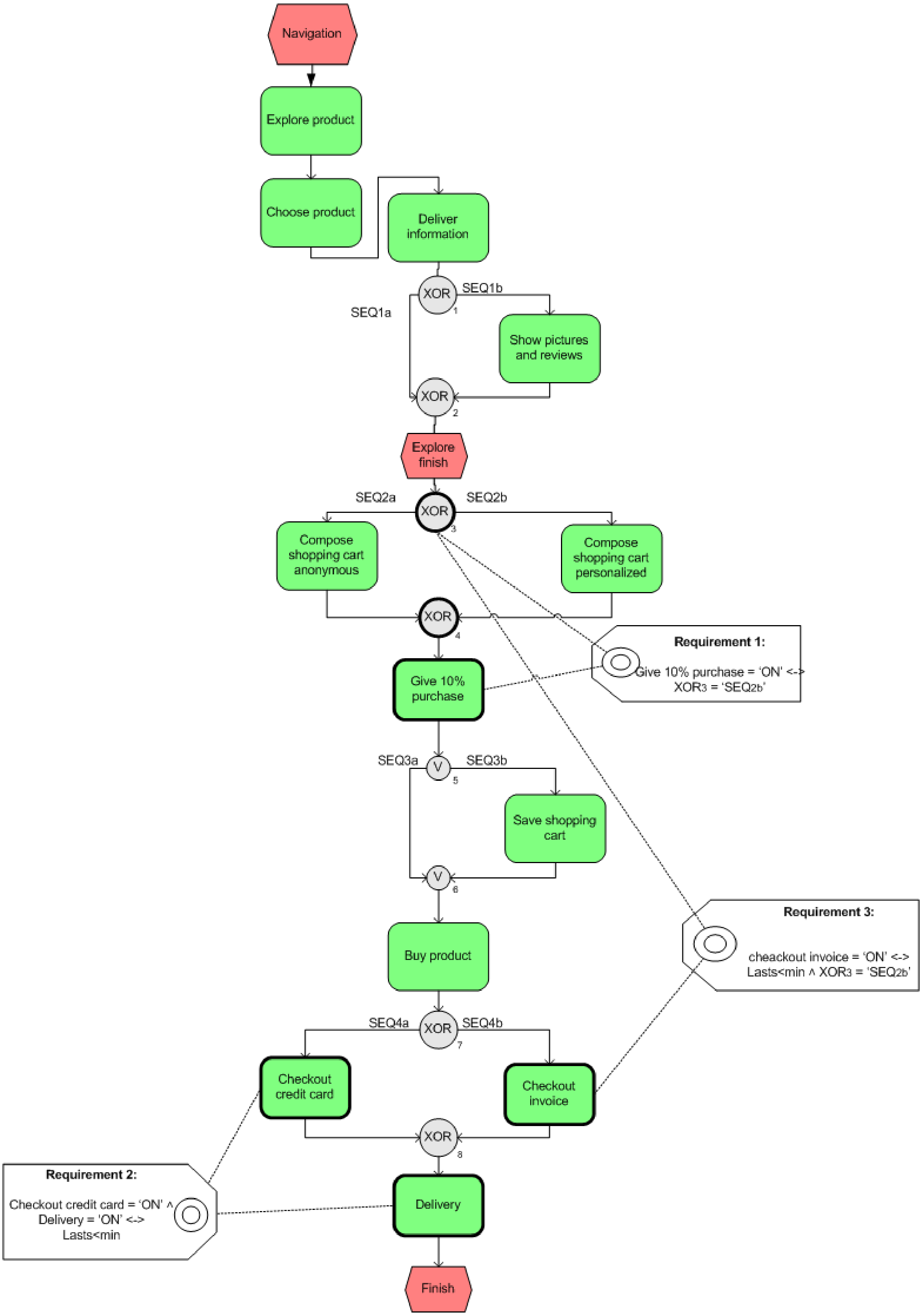

- **Approach: Rich BPMN (PESOA)**

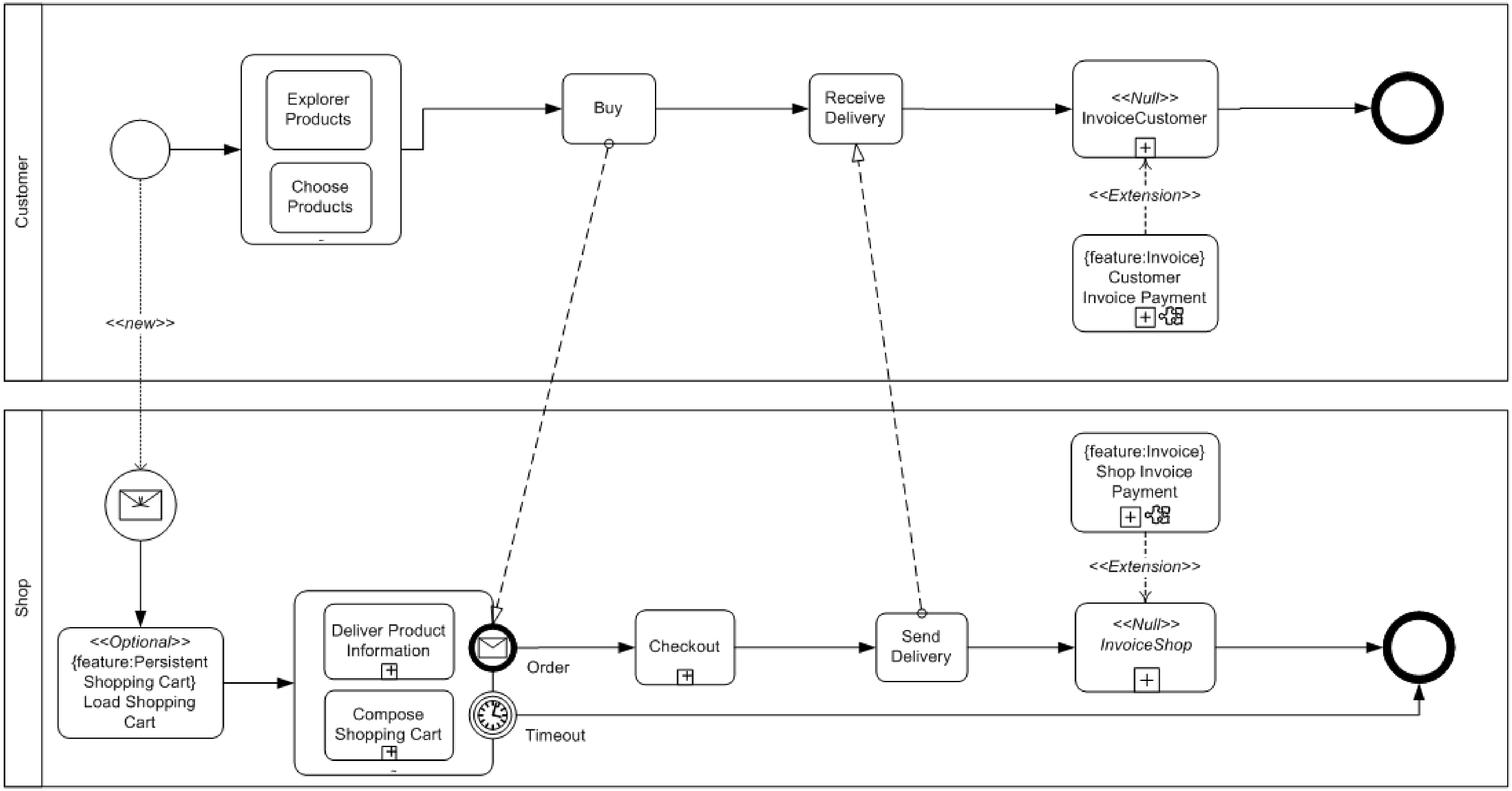

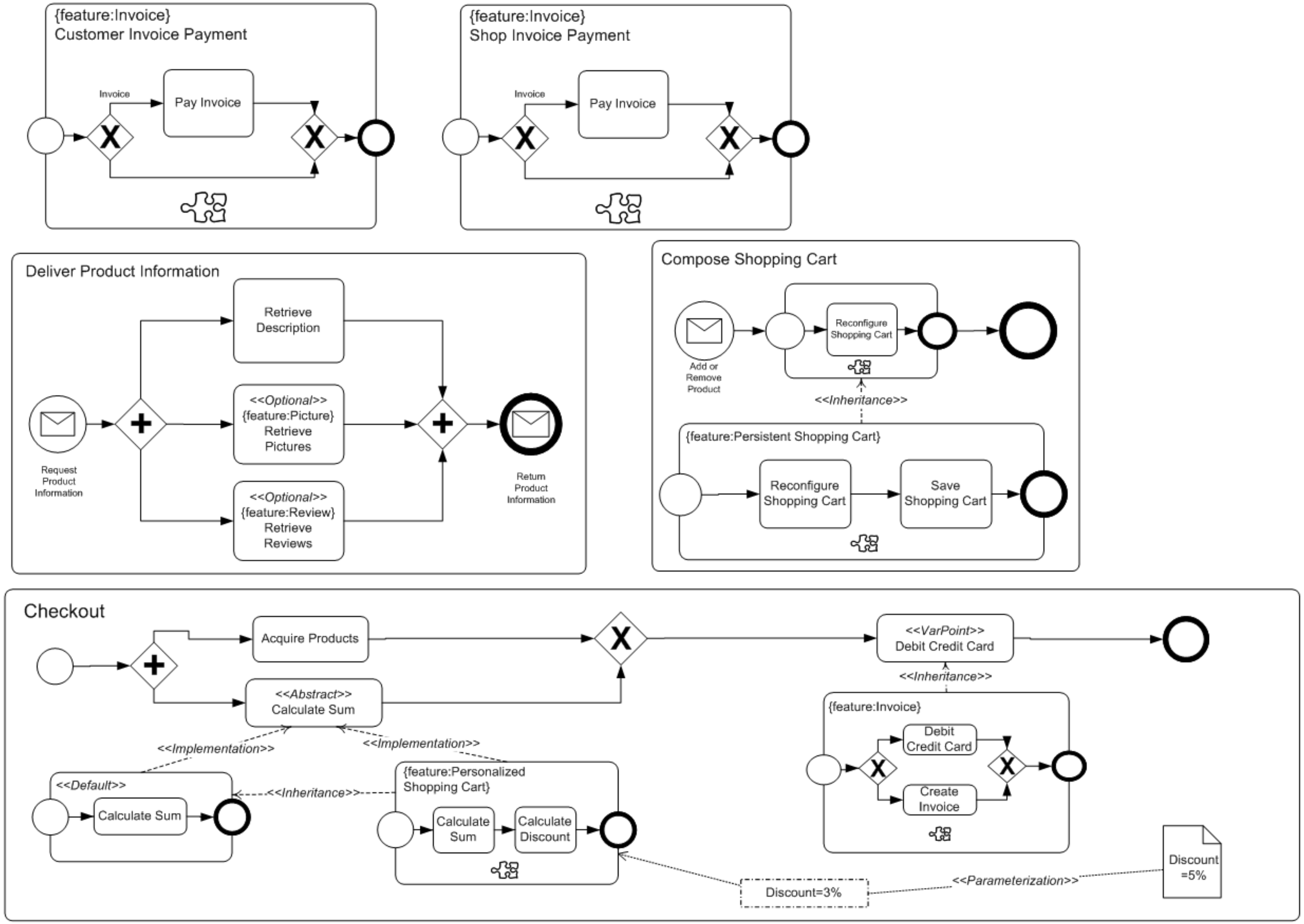

- **Approach: Process Variants by Options (PROVOP)**

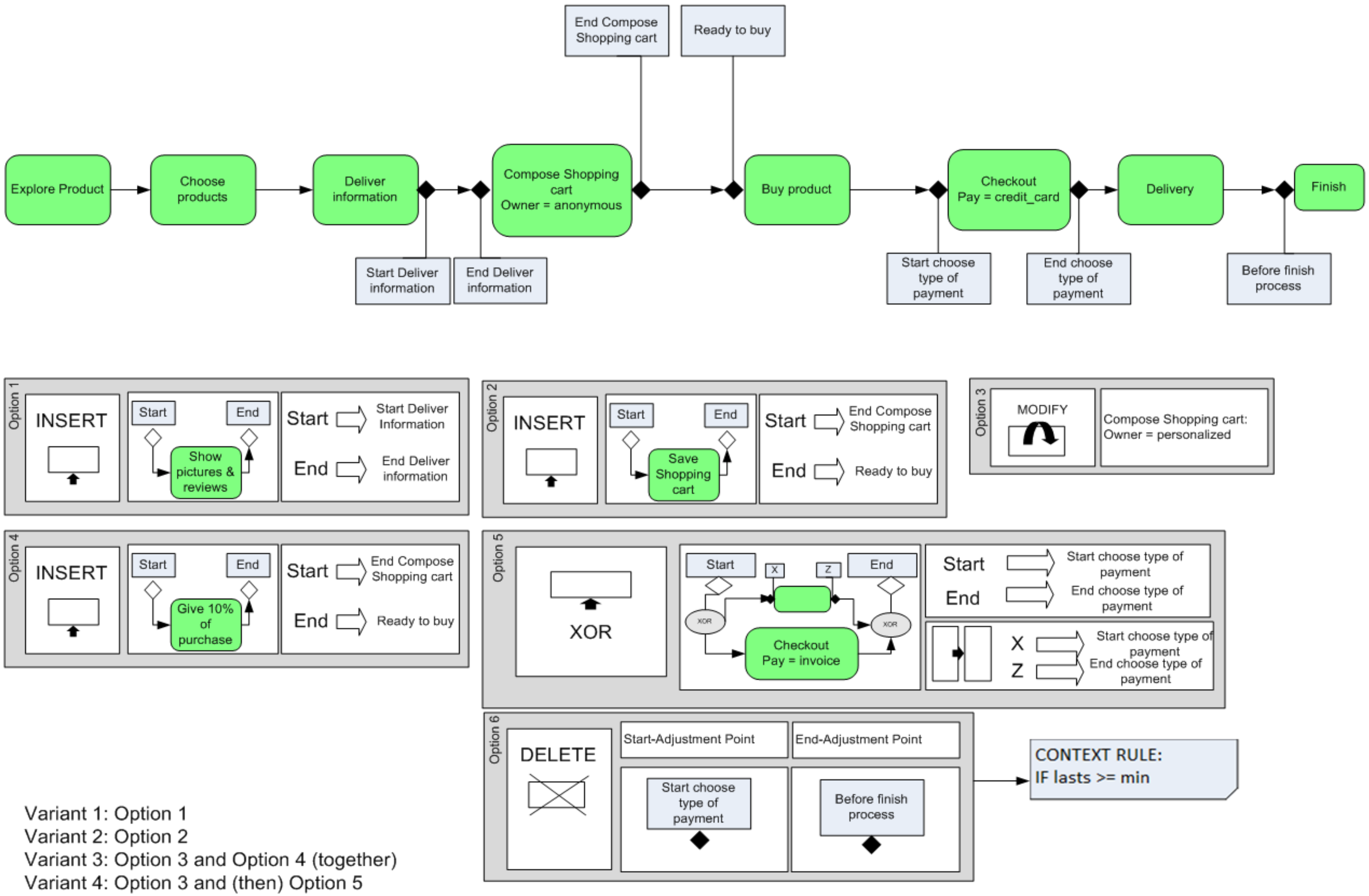

Variant 1: Option 1
Variant 2: Option 2
Variant 3: Option 3 and Option 4 (together)
Variant 4: Option 3 and (then) Option 5

- **Approach: Worklets (Yawl+RDR)**

Worklet-Enabled process:

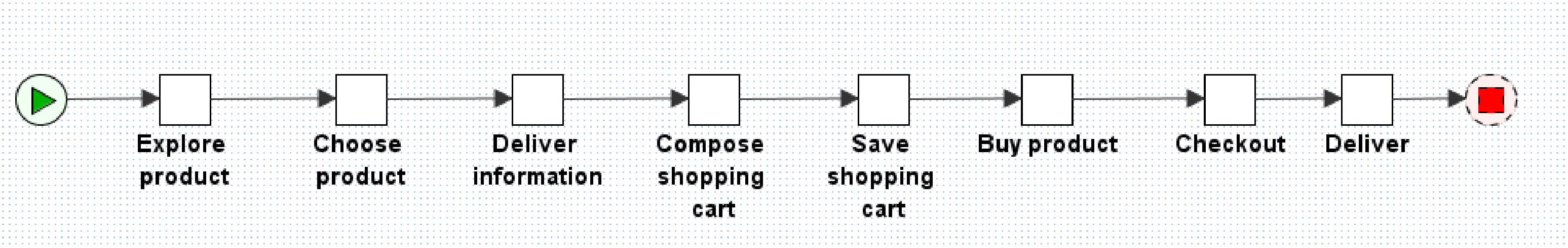

Selection Rule Tree for the "Deliver information" task:

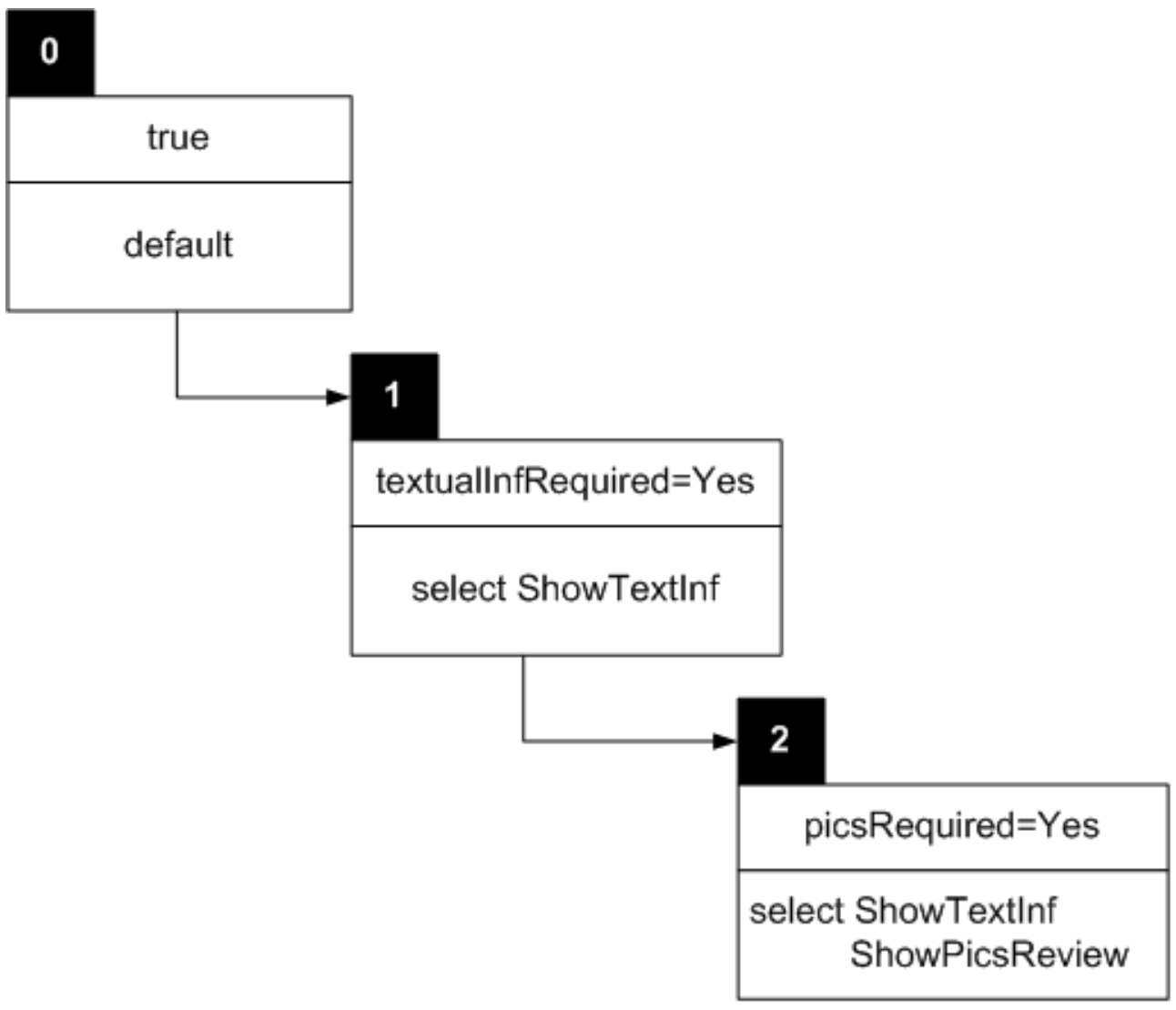

Selection Rule Tree for the "Compose shopping cart" task:

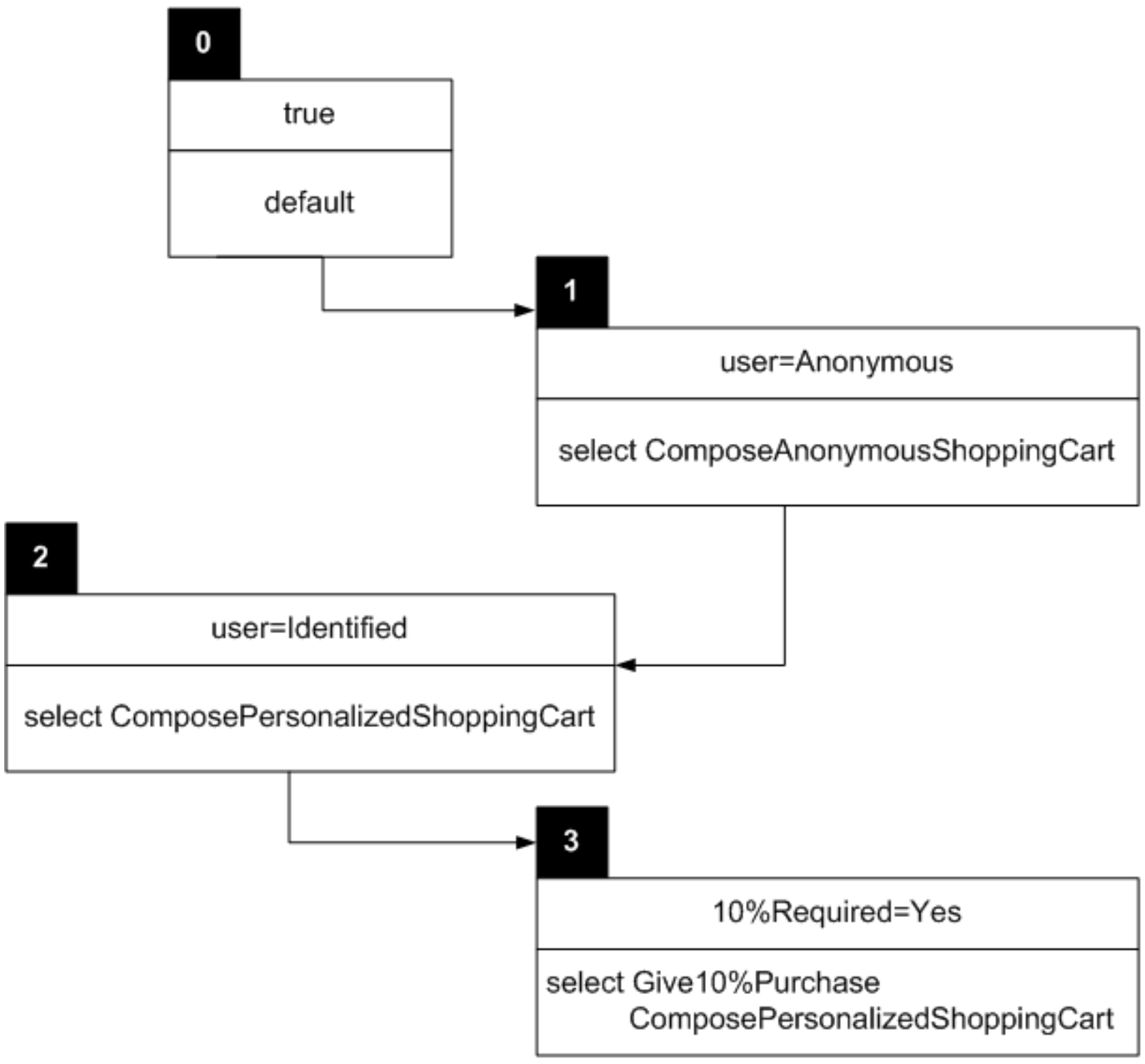

Selection Rule Tree for the “Save shopping cart” task:

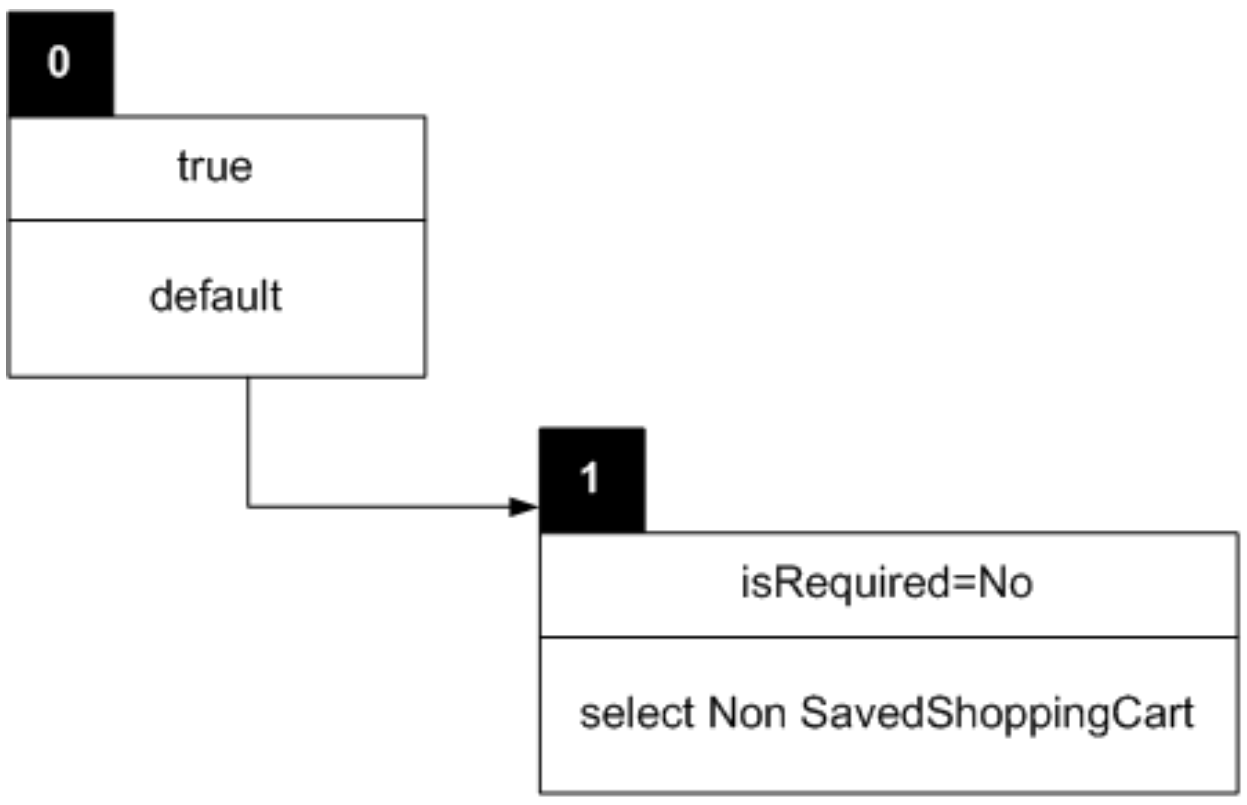

Selection Rule Tree for the “Checkout” task:

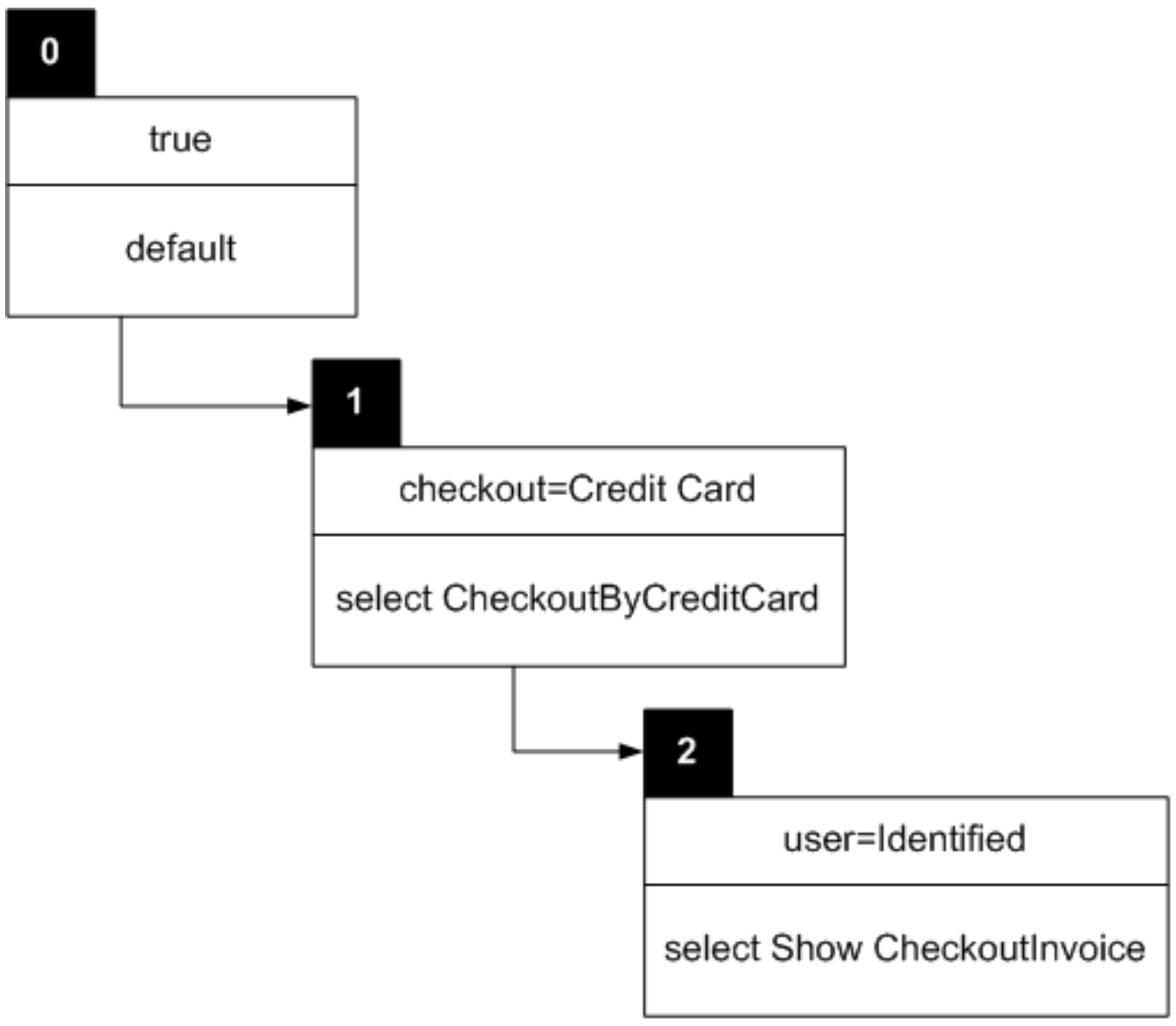

Selection Rule Tree for the “Deliver” task:

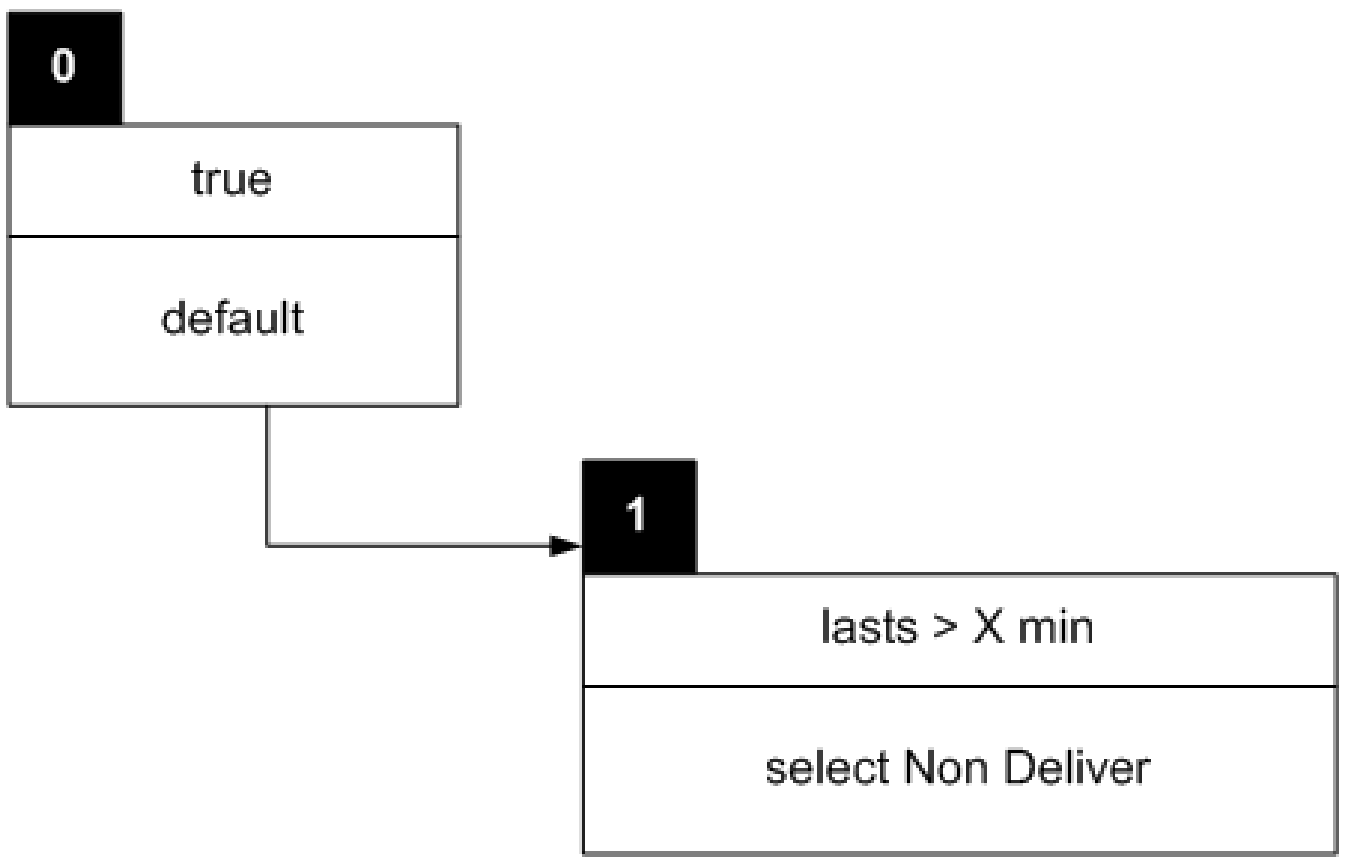

Show TextInf:

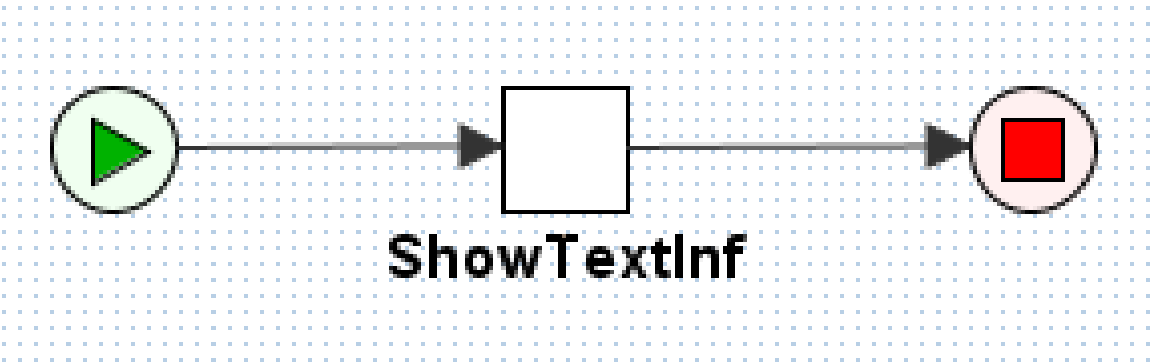

Show PicsReview:

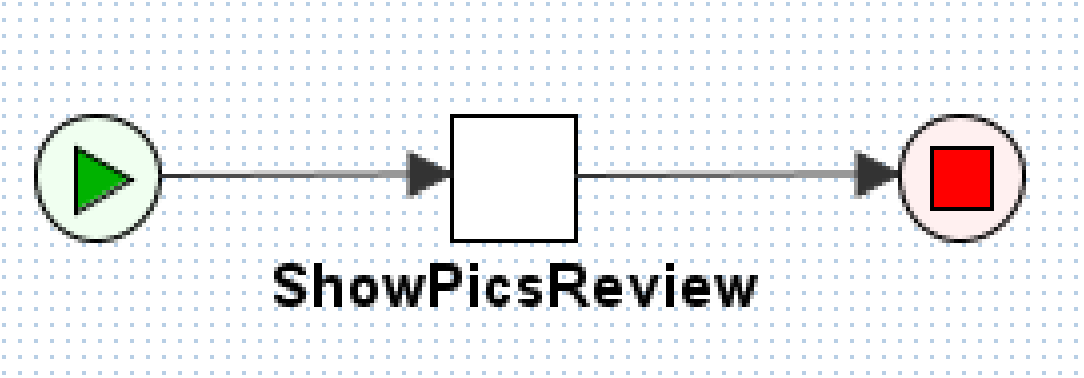

Compose AnonymousShoppingCart:

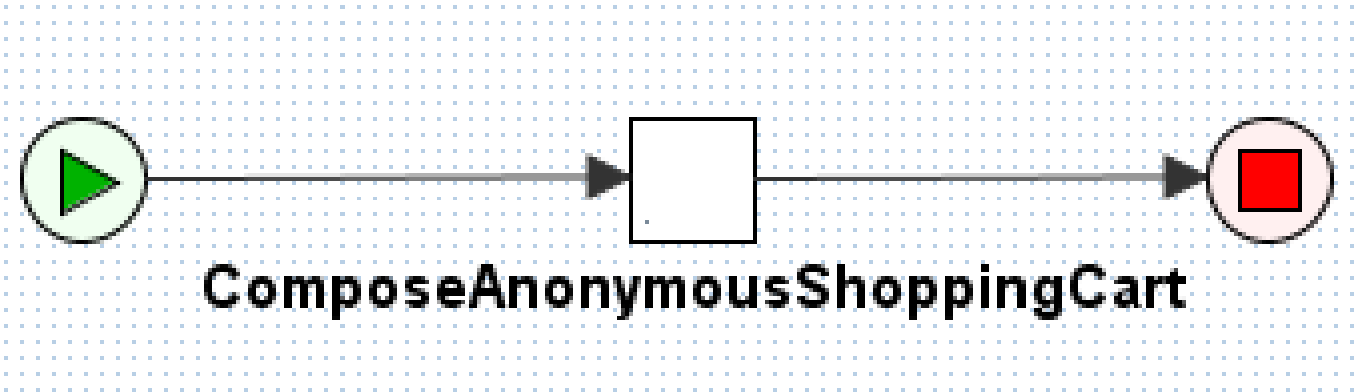

ComposePersonalizedShoppingCart:

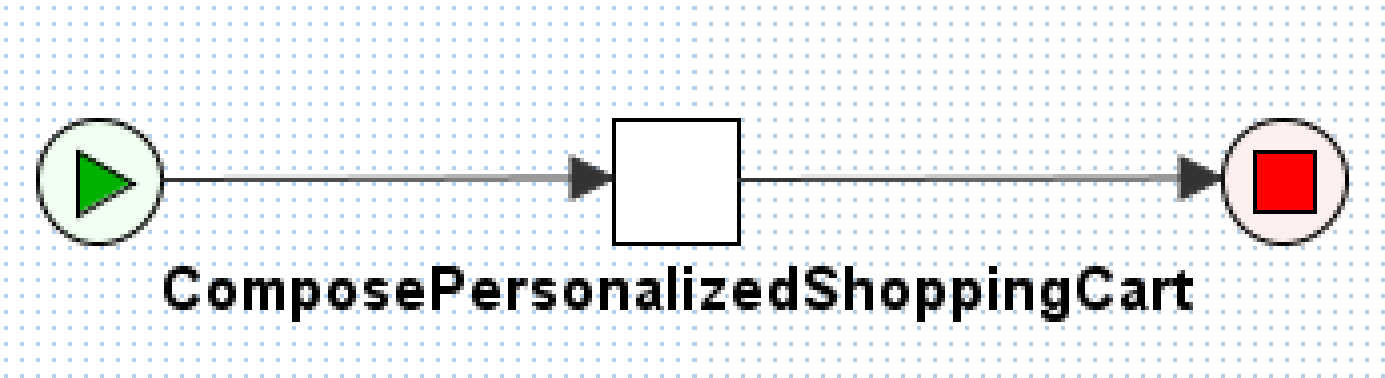

Give10%Purchase:

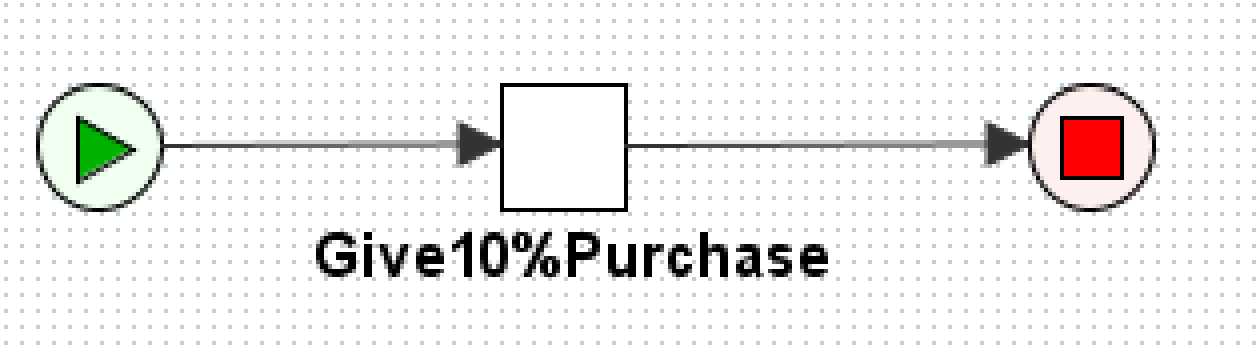

Non SavedShoppingCart:

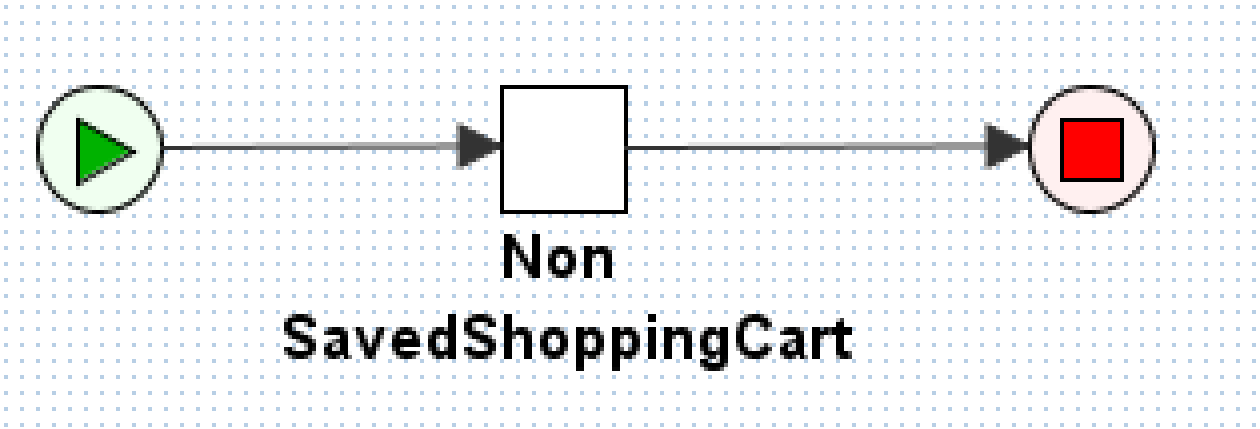

CheckoutByCreditCard:

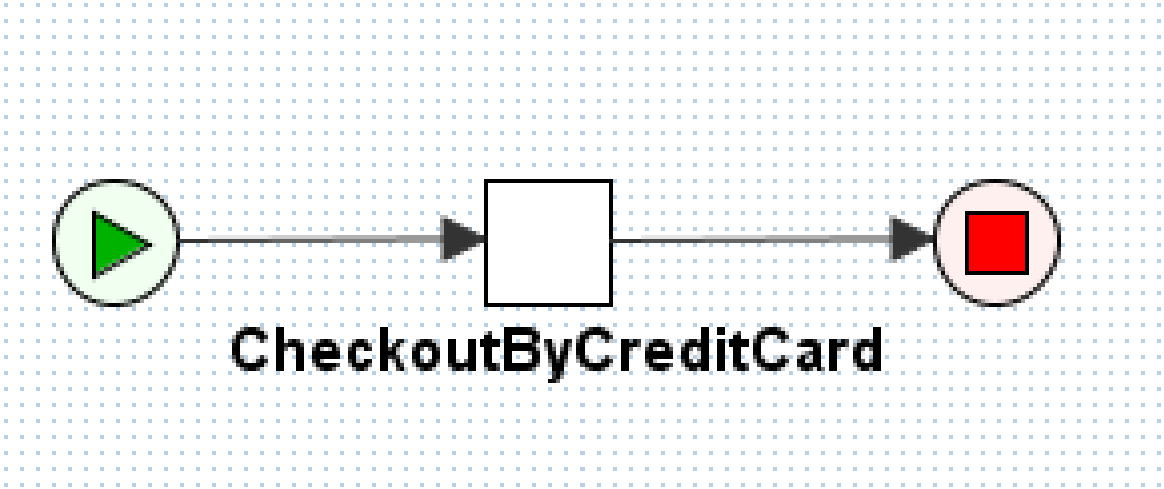

Show CheckoutInvoice:

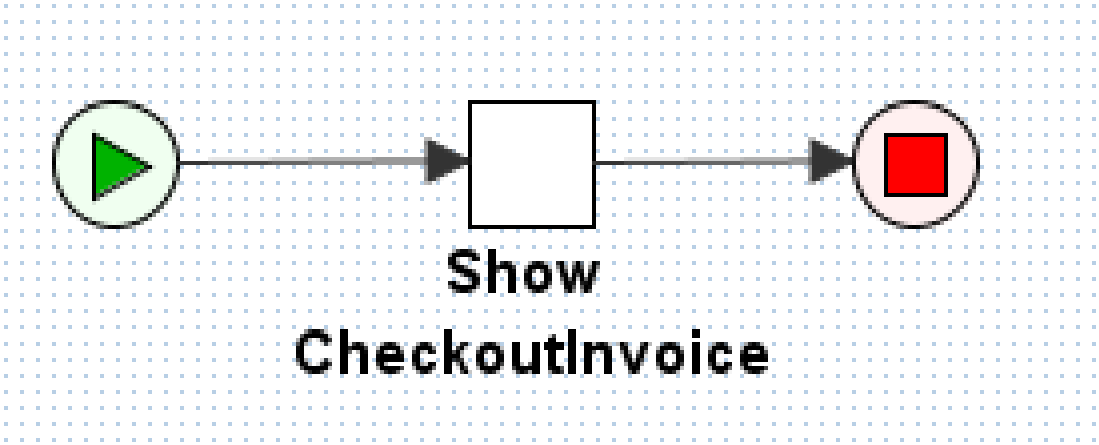

Non Deliver:

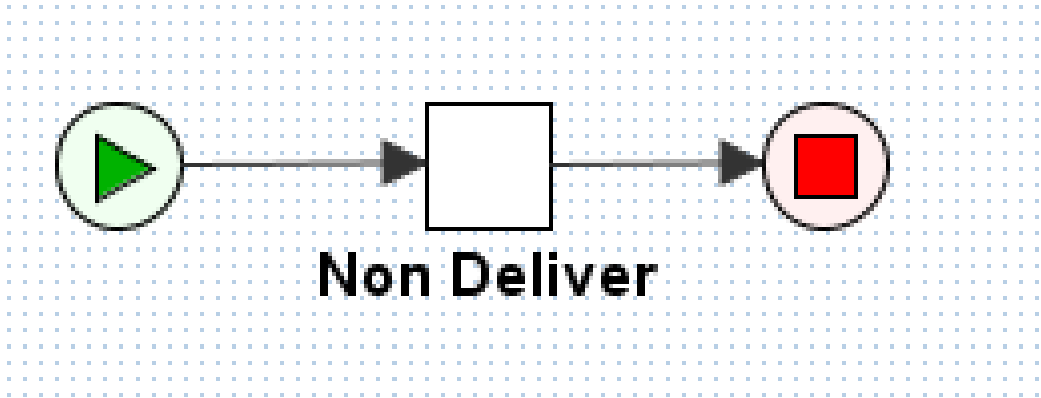

# 4. Conclusions

Conducting these three case studies has helped us, not only to improve our knowledge and expertise of the selected approaches, but also to identify their strengths and weaknesses. The conclusions of our modeling experience is presented below.

### 4.1 Configurable Event-driven Process Chains (C-EPC)

C-EPC is an approach to explicitly capture variability in process models by extended the EPC notation. An easy task when modeling using C-EPC is the identification of those places within the model that may vary, the variation points. However, in C-EPC the resolution of these variation points is attached to other previous decisions of the same model, instead of being based on context aspects. But process modeling is concerned with context related aspects as well, a good example of it can be found in the Variant 5 of the e-business shop example: "if the customer takes a long time to perform the purchase, then the process is finished". C-EPC does not provide any technique to solve this situation, lacking on support context modeling concepts. Thus, it is not possible to model/consider in C-EPC variability that occur only during the execution of the process (run time).

Since C-EPC models are integrated representations, all the process variants are defined together within the model. In order to being able to combine all of them in one unique model, modelers need to have a clear prior idea of the entire process, which is very difficult in models that include a high level of variability, e.g. e-business shop. As a consequence, the models tend to get big and complex very fast which implies that its simplicity disappears.

To restrict variant combination is necessary to define configuration requirements and guidelines using logical predicates, but C-EPC does not provide any technique to check the consistency of them. Thus, it needs to be done manually, which is very time-consuming.

### 4.2 Rich BPMN (PESOA)

The aim of the approach is to improve re-usability and customization of those systems that are developed from the specification of process models. For such purpose and taking into account variability aspects, different stereotypes have been defined in order to cover different variant behaviors. Despite this, the main problem of the PESOA approach is that it is not specified how variation points should be solved. When modelers detect the places that the process may vary, how they should transform them into to those stereotypes that suit better for them is not clearly defined. This leaves the variant definition for modeler interpretation which may lead to correctness issues in the resulting models, requiring to the modelers a high level of expertise to use this approach.

On the contrary, a positive aspect of the PESOA approach is the combination of feature diagrams and BPMN. This combination facilitates the configuration of process models by capturing context aspects.

### 4.3 Process Variant by Options (PROVOP)

Provop provides an operational approach for managing process variants. In particular, the process variants can be configured by applying a set of high-level change operations to a common base process. The introduction of these change operations allows modelers to define the common process separately from individualities as well as to distinguish them, making the models simpler and more intuitive. A good example of that is the resulting model of the case study e-business shop where, despite having a high level of variability, the model is only one workflow in which tasks may be added, delete or modify if necessary.

Provop also provides support for context-aware process configuration by means of the context rules. Nevertheless this rules do not explicitly specify the time in which variation points are solved (design time or run time), which is important in order to configure the process before the deployment.

### 4.4 Worklets (Yawl+RDR)

The Worklet approach enables late binding of process fragments to process activities at run time. Thus, at design time, the activity is merely modeled as placeholder and, at run time, the appropriate process fragment is selected to bound to the process activities. This late binding allows a better activity re-using as well as model understanding. The problem appears in how activities that are going to be bounded (variation points) are identified within the model. There is not provided any special mark to distinguish commonalities from individualities.

Another drawback identified is how to model these activities that may be skipped during the execution. For instance, the MRT activity of the health care process case study should be skipped if the patient has a pacemaker in his/her heart. How to model this pacemaker condition is not clearly specified in the related literature. Moreover, how the activity should be replaced in order to be skipped (i.e. should it be replaced by an empty activity?) or how the RDR should define this replacement are neither clarify in the documentation.